%% file: main.tex
\pdfoutput=1

\documentclass{elsarticle}

\usepackage[latin1]{inputenc}
\usepackage{amsmath}
\usepackage{amssymb}
\usepackage{amsthm}
\usepackage{colortbl}
\usepackage{listings}
\usepackage{enumerate}
\usepackage{textcomp}
\usepackage{times}
\usepackage{graphicx}
\usepackage{minitoc}
\usepackage{breakcites}
\usepackage[title,titletoc]{appendix}
\usepackage{pdfpages}
\usepackage{mdframed}
\usepackage{import}
\usepackage{breakurl}
\usepackage{sidenotes}
\usepackage[a4paper,pagebackref,breaklinks,hidelinks,hyperindex=true]{hyperref}
\usepackage[usenames,dvipsnames]{xcolor}
\usepackage{enumerate}
\usepackage{dashrule}
\usepackage{multirow}
\usepackage[all,cmtip]{xy}
\usepackage{tikz}
\usepackage{pgfplots}
\usetikzlibrary{arrows}
\usetikzlibrary{positioning}

\input{macros}

\input{macros_compat}
\input{macros_jolie_web}

\usepackage{lipsum}
\makeatletter
\def\ps@pprintTitle{%
 \let\@oddhead\@empty
 \let\@evenhead\@empty
 \def\@oddfoot{}%
 \let\@evenfoot\@oddfoot}
\makeatother

\begin{document}

\title{Process-aware web programming with Jolie}

\author{Fabrizio Montesi}
\ead{fmontesi@imada.sdu.dk}
\address{Department of Mathematics and Computer Science,
University of Southern Denmark,
Campusvej 55, 5230 Odense M, Denmark.
Phone number: +4565507171
}

\input{abstract}


\maketitle

\setJolie

\section{Introduction}\label{sec:intro}
\input{introduction}

\section{Overview of Jolie}\label{sec:jolie}
\input{jolie}

\section{Extending Jolie to HTTP}\label{sec:http}
\input{http}

\section{Web Servers}\label{sec:leonardo}
\input{leonardo}

\section{Sessions}\label{sec:sessions}
\input{sessions}

\section{Layering}\label{sec:architectural}
\input{architectural}

\section{RESTful Services}\label{sec:rest}
\input{rest}

\section{Performance}\label{sec:performance}
\input{performance}

\section{Related Work}\label{sec:related}
\input{related}

\section{Discussion and Future Extensions}
\label{sec:discussion}
\label{sec:future}
\input{discussion}

\section{Conclusions}\label{sec:conclusions}
\input{conclusions}

\section*{Acknowledgements}
The author thanks Claudio Guidi, Saverio Giallorenzo, and the anonymous
referees for their useful comments.
This work was supported by CRC (Choreographies for Reliable and efficient Communication software), grant no. 
DFF--4005-00304 from the Danish Council 
for Independent Research.

\bibliographystyle{plain}
\bibliography{biblio}

\end{document}

%% file: macros.tex
\newtheorem{theorem}{Theorem}[section]

\newtheorem{example}[theorem]{Example}

\newtheorem{remark}[theorem]{Remark}

\newcommand{\code}[1]{\texttt{#1}}

\newcommand{\sephline}{\vspace{4mm}\\
\arrayrulecolor{gray}\hline
\arrayrulecolor{black}
\\}

\newcommand{\co}[1]{\overline{#1}}

\definecolor{light-gray}{gray}{0.928}

\definecolor{color:keyword}{rgb}{0.53,0.05,0.05}
\definecolor{color:comment}{rgb}{0.25,0.37,0.75}
\definecolor{color:string}{rgb}{0.87,0.0,0.0}
\usepackage{bold-extra}

\lstdefinelanguage{Jolie}{
morekeywords={
	provide,until,OneWay,RequestResponse,new,
	main,define,inputPort,outputPort,init,execution,include,
	cset,if,else,csets,interface,type,throws,global,constants,for,
foreach,while,int,double,raw,void,undefined,string,long,bool,any,single,
sequential,concurrent,Jolie,Java,JavaScript,embedded,Location,Protocol,
Interfaces,Aggregates,scope,install,cH,comp,throw,this,default,synchronized,
nullProcess,false,true
},
sensitive=true,
morecomment=[l]{//},
morecomment=[s]{/*}{*/},
morestring=[b]",
otherkeywords={;,|,:}
}

\lstdefinelanguage{Chor}{
morekeywords={
	main,program,protocol,define,public,ask,local,if,else,start,show,
	string,long,bool,int,void,end
},
sensitive=true,
morecomment=[l]{//},
morecomment=[s]{/*}{*/},
morestring=[b]",
otherkeywords={@,;,->,|,:}
}

\lstdefinelanguage{JavaScript}{
  keywords={typeof, new, true, false, catch, function, return, null, catch, switch, var, if, in, while, do, else, case, break},
  keywordstyle=\color{color:keyword}\bfseries,
  ndkeywords={class, export, boolean, throw, implements, import, this},
  ndkeywordstyle=\color{darkgray}\bfseries,
  identifierstyle=\color{black},
  sensitive=false,
  comment=[l]{//},
  morecomment=[s]{/*}{*/},
  commentstyle=\color{purple}\ttfamily,
  stringstyle=\color{red}\ttfamily,
  morestring=[b]',
  morestring=[b]"
}

\newcommand{\setJavascript}{\lstset{language=JavaScript}}
\newcommand{\setJolie}{\lstset{language=Jolie}}
\newcommand{\setXML}{\lstset{language=XML}}

%% file: macros_compat.tex
\newcommand{\Jolie}{Jolie}

\newcounter{ncomm}

\newcommand{\lstcode}[1]{\lstinline~#1~}

\newcommand{\lblread}[1]{\textsf{read} \ #1}

\newcommand{\genericifthenelse}{\ifthenelse{e}{P}{Q}}

\newcommand{\genChoice}{\lstcode{[} $\eta_1$ \lstcode{] \{ } $B_1$ \lstcode{ \}}
\ldots
\lstcode{[} $\eta_n$ \lstcode{] \{ } $B_n$ \lstcode{ \}}}
\newcommand{\genIfthenelse}{
\lstcode{if(} $e$ \lstcode{)} $B_1$ [\  \lstcode{else}\  $B_2$
 ]
}
\newcommand{\genWhile}{
\lstcode{while(} $e$ \lstcode{)} $B$
}
\newcommand{\genOW}{\lstcode{o(x)}}
\newcommand{\genNoti}{\lstcode{o@OP(}$e$\lstcode{)}}
\newcommand{\genRR}{\lstcode{o(x)(}$e$\lstcode{)\{} $B$ \lstcode{\}}}
\newcommand{\genSR}{\lstcode{o@OP(}$e$\lstcode{)(y)}}
\newcommand{\terminal}[1]{\textsf{#1}}

\newcommand{\csomit}[1]{}

\tikzstyle{wto} = [->>]

\definecolor{light-gray}{gray}{0.88}

\definecolor{cool}{gray}{0.95}


%% file: macros_jolie_web.tex
\lstdefinelanguage{http}{
morekeywords={
    HTTP, GET, POST, PUT, DELETE,
    Content-Type
},
sensitive=true
}

\newcommand{\setHttp}{\lstset{language=http}}
\newcommand{\setHtml}{\lstset{language=HTML}}

\renewcommand{\genChoice}{\lstcode{[} $\eta_1$ \lstcode{] \{ } $B_1$ \lstcode{ 
\}}
\ldots
\lstcode{[} $\eta_n$ \lstcode{] \{ } $B_n$ \lstcode{ \}}}
\renewcommand{\genIfthenelse}{
\lstcode{if(} $e$ \lstcode{)} $B_1$\  \lstcode{else}\  $B_2$
}
\renewcommand{\genWhile}{
\lstcode{while(} $e$ \lstcode{)} $B$
}
\renewcommand{\genOW}{\lstcode{o(x)}}
\renewcommand{\genNoti}{\lstcode{o@OP(}$e$\lstcode{)}}
\renewcommand{\genRR}{\lstcode{o(x)(}$e$\lstcode{)\{} $B$ \lstcode{\}}}
\renewcommand{\genSR}{\lstcode{o@OP(}$e$\lstcode{)(y)}}
\renewcommand{\terminal}[1]{\textsf{#1}}

\renewcommand{\code}[1]{\texttt{#1}}

\renewcommand{\ifthenelse}[3]{\code{if(} #1
\code{)\,\{}#2\code{\}\,else\,\{} #3 \code{\}}}
\renewcommand{\lblread}[1]{\textsf{read} \ v}

\renewcommand{\genericifthenelse}{\ifthenelse{e}{P}{Q}}

%% file: abstract.tex
\begin{abstract}
We extend the Jolie programming language to capture the native modelling of process-aware web information systems, i.e., web information systems based upon the execution of business processes.
Our main contribution is to offer a unifying approach for the programming of distributed architectures on the web, which can capture web servers, stateful process execution, and the composition of services via mediation.
We discuss applications of this approach through a series of examples that cover, e.g., static content serving, multiparty sessions, and the evolution of web systems. Finally, we present a performance evaluation that includes a comparison of Jolie-based web systems to other frameworks and a measurement of its scalability.
%
%
\end{abstract}

%% file: introduction.tex
A Process-Aware Information System (PAIS) is an information system based upon the execution of 
business processes. These systems are required in many application
scenarios, from inter-process communication to
automated business integration~\cite{pais:book}.
Processes are typically expressed as structures that determine
the order in which communications should be performed
in a system. These structures can be complex and of different kinds; a systematic account 
can be found at~\cite{wfpatterns:website}.
For this reason, many formal methods~\cite{DBLP:conf/apn/Aalst97,G07,LPT07,CM13},
tools~\cite{yawl:journal,HKPYH10,GVRGC10,HMBCY11,MGZ14}, and
standards~\cite{bpel,wscdl,bpmn} have been developed to provide languages for
the definition, verification, and execution of processes.
In these works, compositionality plays a key role to make the development of processes manageable.
For example, in approaches based on process calculi,
complex process structures are obtained by composing simpler ones through the usage of standard
composition operators such as sequence, choice, and parallel (see, e.g.,~\cite{LM07}).
Other approaches follow similar ideas using graphical formal models, e.g.,
Petri Nets~\cite{A98,P77}.
%

In the last two decades, web applications have become increasingly
process-aware.
Web processes -- i.e., processes inside of a web information system --
are usually implemented server-side on top of \emph{sessions},
which track incoming messages related to the same conversation.
Sessions are supported by a shared memory space that lives through different client invocations.
Differently from the aforementioned approaches for designing processes, the
major languages and platforms for developing web applications
(e.g., PHP, Ruby on Rails, and Java EE) do not support the explicit
programming of process structures.
As a workaround, programmers have to simulate processes using bookkeeping variables in
the shared memory space of sessions.
For example, consider a process in a Research Information Service (RIS)
where a user has to authenticate through a
\code{login} operation before accessing another
operation, say \code{addPub}, for registering a publication.
This would be implemented by defining the 
\code{login} and the \code{addPub} operations separately. The
code for \code{login} would update a bookkeeping variable in the session
state and the implementation for \code{addPub} would check that variable
when it is invoked by the user.
Although this approach is widely adopted, it is also error-prone: since
processes can assume quite complex structures, simulating them
through bookkeeping variables soon becomes cumbersome.
Consequently, the produced code may be poorly readable and hard to
maintain.

The limitations described above can be avoided by adopting a multi-layered
architecture.
For example, it is possible to stratify an application by employing: a web
server technology (e.g., Apache Tomcat) for serving content to web browsers; a
web scripting language (e.g., PHP) for programmable request processing;
a process-oriented language (e.g., WS-BPEL~\cite{bpel}) for modelling the
application processes; and, finally, mediation technologies such as proxies and
ESB~\cite{esb:book} for integrating the web application within larger systems.
Such an architecture would offer a good separation of concerns.
However, the resulting system would be highly heterogeneous, requiring a
specific know-how for handling each part. Thus, it would be hard to maintain
and potentially prone to breakage in case of modifications.

The aim of this paper is to simplify the programming of
process-aware web information systems. We build our results on top 
of \Jolie{}, a general-purpose service-oriented programming language that can handle both the
structured modelling of processes and their
integration within larger distributed systems~\cite{MGZ14,jolie:website}.
Jolie is briefly introduced in \S~\ref{sec:jolie}.

\subsection{Contributions}
%
Our main contribution is the extension of Jolie to obtain a unifying technology for the programming of processes, web technologies (web servers and scripting) and mediation services (e.g., proxies), which facilitates the development of heterogeneous information systems that involve the Web. We then investigate the applicability of this framework, showing that our extended version of Jolie captures the different components of web systems and their integration using a homogeneous set of concepts.
We proceed as described below.

\paragraph{Web processes}
We extend the \Jolie{} interpreter to support the HTTP protocol,
enabling processes written in \Jolie{} to send and receive HTTP messages
(\S~\ref{sec:http}).
The integration is \emph{seamless}, meaning that the processes defined in \Jolie{} remain abstract from the underlying
data formats used in the Web: data structures in Jolie are transparently
transformed to HTTP message payloads and vice versa (\S~\ref{sec:transform}).
These transformations can be configured using \emph{port parameters}, an extension of the Jolie language that we develop to allow processes to map information from HTTP headers to application data and vice versa (\S~\ref{sec:config}). Parameters support mobility: they can be transparently transmitted from service registries to clients, allowing clients to be transparently configured at runtime with the right binding information (\S~\ref{sec:http_examples}, Example~\ref{ex:param_mobility}).

\paragraph{Web servers as processes}
We develop a web server, called Leonardo (\S~\ref{sec:leonardo}), using our approach.
The web server is given as a simple process that
(i) receives the name of the resource a client wants to access, then
(ii) reads the content of such resource,
and (iii) sends the content back to the client.
%
%
%
Leonardo is an example of the fact that, in our framework, a web server is not a separate
technology but it is instead a simple case of process.
We also show how to extend Leonardo to handle simple CRUD operations over HTTP.

\paragraph{Sessions}
We combine our HTTP extension for Jolie with message correlation, a mechanism used in
service-oriented technologies to route incoming messages
to their respective processes running inside of a service~\cite{bpel,MGZ14}.
We first show that this combination is adequate wrt existing practice:
it enables Jolie processes to use the standard methodology of tracking
client-server web sessions using unique session identifiers (\S~\ref{sec:binary}).
Then, we generalise such methodology to program multiparty sessions,
i.e., structured conversations among a process and
multiple external participants~\cite{HYC08} (\S~\ref{sec:multiparty}).

\paragraph{Architectural programming}
We present how to obtain separation of concerns in a web architecture
implemented with our approach, by combining HTTP with aggregation,
a Jolie primitive for programming the structure of service
networks~\cite{M10,DGGMM12,MGZ14} (\S~\ref{sec:architectural}).
We demonstrate the usefulness of this combination by implementing a
multi-layered system that integrates different components.
We also discuss how to deal with the evolution of software
architectures obtained with our approach (\S~\ref{sec:evolvability}).
%

\paragraph{RESTful services}
We discuss how to develop web systems based on the REST style~\cite{F00} with our framework, using URI templates to bridge resource-oriented interactions to processes (\S~\ref{sec:rest}).
The standard separation of concerns between routing and business logic can be achieved by developing a routing service that routes REST requests to other services, which required developing a reflection library for the Jolie language (\S~\ref{sec:rest_router}).
We then show how structured processes can be combined with our REST router to obtain RESTful multiparty sessions (\S~\ref{sec:rest_processes}).
Since, in these scenarios, requests typically have to be validated both at client- and server-side, we provide an integration between Jolie and JavaScript to be able to run the same validation code, achieving the same de-duplication benefits as in frameworks based on JavaScript (\S~\ref{sec:javascript}).

\paragraph{Performance}
A performance evaluation of our approach is given (\S~\ref{sec:performance}). This evaluation covers two main aspects. First, our data shows that our solution has comparable performance to that of other existing frameworks in the execution of basic tasks, such as offering static files or templated web pages. Second, we analyse the scalability of our approach wrt the number of services involved in the computation of responses to clients.

%% file: jolie.tex
\Jolie{}~\cite{MGZ07} is a general-purpose service-oriented
programming language, released as an open-source project~\cite{jolie:website}
and formally specified as a process calculus~\cite{G07,MC11}.
In this section, we briefly describe some aspects of Jolie that are
relevant for our discussion.
We refer the interested reader to~\cite{MGZ14} and~\cite{jolie:website} for a more comprehensive
presentation of the language. Readers who are already familiar with the Jolie language may
skip this section and resume reading from \S~\ref{sec:http}.

\subsection{Jolie programs}
\label{sec:jolie_programs}
Every \Jolie{} program defines a service
and consists of two parts: \emph{behaviour} and \emph{deployment}.
A behaviour defines the implementation of the operations offered
by the service; it consists of communication and computation instructions,
composed into a structured process (a workflow) using constructs such as
sequences, parallels, and internal/external choices.
Behaviours rely on \emph{communication ports} to perform communications, which
are to be defined in the deployment part.
The latter can also make use of architectural primitives for handling
the structure of an information system.
Formally, a \Jolie{} program is structured as: 
\begin{center}
$D$ \quad \lstcode{main \{ } $B$ \lstcode{\}}
\end{center}
Above, $D$ represents the deployment and $B$ the behavior of the program.

\subsection{Behaviour}
\label{sec:behaviour}
We report (a selection of) the syntax of behaviours in Figure~\ref{fig:web:jolie_behaviour}.
A behaviour $B$ can use primitives for performing communications, computation,
and their composition in processes. We briefly comment the syntax.
%
\input{jolie_behavioural_syntax}
Terms \emph{(input)}, \emph{(output)},
and \emph{(input choice)} implement communications. An input $\eta$ can either
be a one-way or a request-response, following the WSDL standard~\cite{wsdl}.
Statement \emph{(one-way)} receives a message for operation \lstcode{o} and
stores its content in variable \lstcode{x}.
Term \emph{(request-response)} receives a message for operation \lstcode{o} in
variable \lstcode{x}, executes behaviour $B$ (called the \emph{body} of the
request-response input), and then sends the value of the evaluation of
expression $e$ to the invoker.
Dual to input statements, an output $\co\eta$ can be a 
\emph{(notification)} or a \emph{(solicit-response)}.
A \emph{(notification)} sends a message to \lstcode{OP} containing the value of the evaluation of
expression $e$. Term \emph{(solicit-response)} sends a message to \lstcode{OP} containing the
evaluation of $e$ and then waits for a response from the invoked
service, storing it afterwards in variable \lstcode{y}.
In both \emph{(notification)} and \emph{(solicit-response)},
\lstcode{OP} is the name of an \emph{output port}, which acts as a reference
to an external service. Output ports are concretely defined in the deployment part
of a program; we will present them in \S~\ref{sec:deployment}.
%

Term \emph{(input choice)} implements an input-guarded choice; it is similar
to the \lstcode{pick} primitive in WS-BPEL~\cite{bpel}.
Specifically, the construct waits for a message for any of the inputs
in $\eta_1,\ldots,\eta_n$. When a message for one of these inputs is received, 
say for $\eta_i$ where
$1\leq i \leq n$, then the statement is executed as follows, in order: (i) all the other branches in the choice
(i.e., all the
\lstcode{[} $\eta_j$ \lstcode{] \{ } $B_j$ \lstcode{ 
\}} such that $j\neq i$) are discarded; (ii) $\eta_i$ is executed;
and, finally, $B_i$ is executed.

Terms \emph{(cond)} and \emph{(while)} implement, respectively, the standard
conditional and iteration constructs.
Term \emph{(seq)} models sequential execution and reads as: execute $B$,
wait for its termination, and then run $B'$.
In term \emph{(par)}, instead, $B$ and $B'$ are run in parallel.
Term \emph{(throw)} throws a fault signal \lstcode{f}, interrupting execution.
If a fault signal is thrown from inside a request-response body, the
invoker of the request-response statement is
automatically notified of the fault~\cite{MGLZ08}.
We omit the syntax for handling faults, which is not necessary for reading
this paper.

Term \emph{(assign)} stores the result of the evaluation of expression $e$ in
variable \lstcode{x}.
Term \emph{(alias)} makes variable \lstcode{x} an alias for
variable \lstcode{y}, i.e., after its execution accessing \lstcode{x} will be equivalent to
accessing \lstcode{y}.
Term \emph{(inact)} denotes the empty behaviour (no-op).

\begin{example}[Structured data]
\Jolie{} natively supports the manipulation of structured data. In
\Jolie{}'s memory model the program state is a tree (possibly with arrays as nodes, 
see~\cite{M10}), and 
every variable, say \lstcode{x}, can be a \emph{path} to a node in the
memory tree. Paths are constructed through the dot ``.'' operator; for instance, the
following sequence of assignments
\begin{lstlisting}[label={listing:person}]
person.name = "John"; person.age = 42
\end{lstlisting}
would lead to a state containing a tree with root label \lstcode{person}.
For the reader familiar with XML, a corresponding XML representation would be:
\setXML
\begin{lstlisting}
<person> <name>John</name> <age>42</age> </person>
\end{lstlisting}
\setJolie
\end{example}

\subsection{Deployment}
\label{sec:deployment}

We introduce now (a selection of) the syntax of deployments.
A deployment includes definitions of \emph{input ports}, denoted by $IP$, and
\emph{output ports}, denoted by $OP$,
which respectively support input and output communications with other services.
Input and output ports are one the dual concept of the other, and their respective syntaxes are quite similar.
Both kinds of ports are based on the three basic elements of \emph{location},
\emph{protocol} and \emph{interface}.
Their syntax is reported in Figure~\ref{fig:web:ports}.
\begin{figure}
\begin{center}
\begin{tabular}{rcl}
\multicolumn{3}{l}{$IP$  ::=  \lstcode{inputPort} $Port$ \qquad
$OP$  ::=  \lstcode{outputPort} $Port$}\\[1mm]
$Port$ & ::= & \terminal{id} \lstcode{\{} \\
& & \lstcode{Location:} $Loc$ \\
& & \lstcode{Protocol:} $Proto$ \\
& & \lstcode{Interfaces:} \terminal{iface}$_1$, \ldots, \terminal{iface}$_n$ \\
& & \lstcode{\}}\\
\end{tabular}
\end{center}
\caption{Jolie, syntax of ports (selection).}
\label{fig:web:ports}
\end{figure}
In the syntax of ports, i.e., term $Port$, $Loc$ is a URI (Uniform Resource Identifier)
that defines the location of the port; $Proto$ is an identifier referring to the 
data protocol to use in the port, which specifies how input or output messages through
the port should be respectively decoded or encoded; the identifiers \terminal{iface}$_1$, \ldots, \terminal{iface}$_n$
are references to the interfaces accessible through the port.

\Jolie{} supports several kinds of locations and protocols.
For instance, a valid $Loc$ for accepting TCP/IP connections on TCP port 8000 would be
\lstcode{"socket://localhost:8000"}. Other supported locations are based, respectively, on Unix
sockets, Bluetooth communication channels, and local in-memory channels (channels implemented
using shared memory).
Some supported instances of $Proto$ are \lstcode{sodep}~\cite{jolie:website} (a binary
protocol, optimised for performance),
\lstcode{soap}~\cite{soap}, and \lstcode{xmlrpc}~\cite{xmlrpc}.

The interfaces referred to by a communication port define the operations
that can be accessed through that port.
Each interface defines a set of operations, along with their respective
(i) operation types, defining if an operation is to be used as a one-way or
a request-response, and (ii) types of carried messages.
For example, the following code
\begin{lstlisting}
interface SumIface { RequestResponse: sum(SumT)(int) }
\end{lstlisting}
defines an interface named \lstcode{SumIface} with a request-response operation,
called \lstcode{sum}, that expects input messages of type \lstcode{SumT} and replies with
messages of type \lstcode{int} (integers).
Data types for messages follow a tree-like structure; for example, we could define
\lstcode{SumT} as follows:
\begin{lstlisting}
type SumT:void { .x:int .y:int }
\end{lstlisting}
We read the code above as: a message of type \lstcode{SumT} is a tree with
an empty root node (\lstcode{void}) and two subnodes, \lstcode{x} and
\lstcode{y}, that have both type \lstcode{int}.

\begin{example}[A complete Jolie program]
\label{example:sum}
We give an example of how to combine behaviour and deployment
definitions, by showing a simple service defined in \Jolie{}.
The code follows:
\begin{lstlisting}[label={lst:sum}]
type SumT:void { .x:int .y:int }

interface SumIface { RequestResponse: sum(SumT)(int) }

inputPort SumInput {
Location: "socket://localhost:8000"
Protocol: soap
Interfaces: SumIface
}

main
{
  sum( req )( resp ) {
    resp = req.x + req.y
  }
}
\end{lstlisting}
Above, input port \lstcode{MyInput} deploys the interface \lstcode{SumIface}
(and thus the \lstcode{sum} operation) on TCP port 8000, waiting for TCP/IP
socket connections by invokers using the \lstcode{soap} protocol.
The behaviour of the service is contained in the \lstcode{main} procedure, the entry
point of execution in Jolie.
The behaviour in \lstcode{main} defines a request-response input on
operation \lstcode{sum}.
In this paper, we implicitly assume that
all services are deployed with the \lstcode{concurrent} execution modality for
supporting multiple session executions,
from~\cite{M10}.
This means that whenever the first input of the behavioural definition of a
service receives a message from the network, \Jolie{} will spawn a dedicated process with a local
memory state to execute the rest of the behaviour. This process will be equipped
with a local variable state and will proceed in parallel to all the others.
Therefore, in our example, whenever our service receives a request for
operation \lstcode{sum} it will spawn a new parallel process instance. The
latter will enter into the body of \lstcode{sum}, assign to variable
\lstcode{resp} the result of adding the subnodes \lstcode{x} and \lstcode{y} of
the request message \lstcode{req}, and finally send back this result to the original invoker.
\qed
\end{example}

%% file: jolie_behavioural_syntax.tex
\begin{figure}
\begin{center}
\begin{tabular}{lclll}
$B$ & ::= & $\eta$ & \quad \emph{(input)} \\
& $|$ & $\co\eta$ & \quad \emph{(output)} \\
& $|$ & \genChoice & \quad \emph{(input choice)}\\
& $|$ & \genIfthenelse & \quad \emph{(cond)}\\
& $|$ & \genWhile & \quad \emph{(while)}\\
& $|$ & $B$ \lstcode{;} $B'$ & \quad \emph{(seq)}\\
& $|$ & $B$ \lstcode{|} $B'$ & \quad \emph{(par)}\\
& $|$ & \lstcode{throw(f)} & \quad \emph{(throw)}\\ 
& $|$ & \lstcode{x = } $e$ & \quad \emph{(assign)}\\
& $|$ & \lstcode{x -> y} & \quad \emph{(alias)}\\
& $|$ & \lstcode{nullProcess} & \quad \emph{(inact)}
\sephline
$\eta$ & ::= & \genOW & \quad \emph{(one-way)} \\
& $|$ & \genRR & \quad \emph{(request-response)} \\
\\
$\co\eta$ & ::= & \genNoti & \quad \emph{(notification)} \\
& $|$ & \genSR & \quad \emph{(solicit-response)} \\
\end{tabular}\\
\end{center}
\caption{\Jolie{}, syntax of behaviours
(selection).}\label{fig:web:jolie_behaviour}
\end{figure}

%% file: http.tex
We extend \Jolie{} to support web applications by introducing a new protocol
for communication ports, named \lstcode{http}, and by extending the language 
of deployments to support configuration parameters for protocols.
The protocol follows the specifications of HTTP, and integrates the
message semantics of Jolie to that of HTTP and its different content encodings.
In this section, we discuss the main aspects of our implementation.

\subsection{Message transformation}\label{sec:transform}
The central issue to address for integrating \Jolie{} with the HTTP protocol is
establishing how to transform HTTP messages in messages for the input and output
primitives of \Jolie{} and vice versa. Our objective is twofold: on the one hand, we aim at having transparent transformations between data payloads inside of HTTP messages (e.g., XML documents or JSON structures) and Jolie values, so that the programmer does not have to deal with them; on the other hand, we also need to give Jolie programs enough low-level control on HTTP messages such that implementing standard components found in web systems is possible (e.g., web servers or REST routers, as we will discuss respectively in \S~\ref{sec:leonardo} and \S~\ref{sec:rest}).
Hereby we discuss primarily how our implementation manages request messages;
response messages are similarly handled.
The (abstract) structure of a \emph{request message} in HTTP is:
\begin{center}
\begin{tabular}{l}
\setHttp
$Method$ \ $Resource$ \ \lstcode{HTTP/}$Version$ \ $Headers$ \ 
$Body$
\end{tabular}
\end{center}
\setJolie
Above, $Method$ specifies the action that the client intends to perform and can be
picked by a static set of keywords, such as \lstcode{GET}, \lstcode{PUT},
\lstcode{POST}, etc.
Term $Resource$ is a URI path telling which resource the client is requesting.
Term $Version$ is the HTTP protocol version of the message.
The term $Headers$ may include descriptive information on the message $Body$, e.g., the type of
its content (\lstcode{Content-Type}),
or parameters that influence the behaviour of the receiver, e.g., the wish
to keep the underlying connection open for future requests
(\lstcode{Connection: keep-alive}).
Finally, $Body$ contains the content (payload) of the HTTP message.

A \Jolie{} message consists of an operation name (the operation the message is meant for)
and a structured value (the content of the message)~\cite{M10}. Hence,
we need to establish where to read or write these elements in an HTTP
message.
For operation names, we interpret the path part of the $Resource$ URI as the
operation name.
The $Method$ of an HTTP message, instead, is read and written by
\Jolie{} programs through a configuration parameter of our extension, described
later in \S~\ref{sec:config}.
The value of a Jolie message is obtained from $Body$ and the
rest of the $Resource$ URI (query and fragment parts). We use the latter
to decode querystring parameters as \Jolie{} values.

The content of an HTTP message may be encoded in one of different formats.
Our \lstcode{http} extension handles querystrings, form encodings (simple and
multipart), XML, JSON~\cite{json}, and GWT-RPC\footnote{We also developed a companion GWT-RPC client library, called
\lstcode{jolie-gwt}, for a more convenient access to web services written in
\Jolie{} by integrating with the standard GWT development tools.
}~\cite{gwt}.
Programmers can use the \lstcode{format} parameter (\S~\ref{sec:config}) to
control the data format for encoding and decoding messages.
Most of the times, however, this decision is performed automatically via standard HTTP content negotiation and the programmer
does not need to know which format is used (\lstcode{format} is a fallback parameter in the case that the client does not ask for a content type).
As an example of message translation, the HTTP message:
\setHttp
\begin{lstlisting}
GET /sum?x=2&y=3 HTTP/1.1
\end{lstlisting}
\setJolie
would be interpreted as a \Jolie{} message for operation \lstcode{sum}. The
querystring \lstcode{x=2&y=3} would be translated to a structured value with
subnodes \lstcode{x} and \lstcode{y}, containing respectively the strings
\lstcode{"2"} and \lstcode{"3"}.

\subsection{Automatic type casting}
\label{sec:type_casting}
Querystrings and other common message 
formats used in web applications, such as
HTML form encodings, do not carry type information.
Instead, they carry only string representations of values; the information on
the types that these values may have had in the code of the sender (e.g., in Javascript)
is therefore lost.
However, type information is necessary for supporting services such as the
sum service in Example~\ref{example:sum}, which specifically requires its input 
values to be integers.
To handle such cases, we introduce the mechanism of \emph{automatic type
casting}. Automatic type casting reads incoming messages that do not carry type
information (such as querystrings or HTML forms) and tries to cast their content
values to the types expected by the service interface for the message operation.
As an example, consider the querystring \lstcode{x=2&y=3} that we discussed before. Since its
HTTP message is a request for operation \lstcode{sum}, the automatic type
casting mechanism would retrieve the typing for the operation and see that
nodes \lstcode{x} and \lstcode{y} should have type \lstcode{int}. Therefore,
it would try to re-interpret the strings \lstcode{"2"} and \lstcode{"3"} as
integers before giving the message to the behaviour of the \Jolie{} program.
There are cases that type casting may fail to handle; for example,
in \lstcode{x=hello} the string
\lstcode{hello} cannot be cast to
an integer for \lstcode{x}. In such cases, our \lstcode{http} protocol will send
a \lstcode{TypeMismatch} fault back to the invoker with HTTP status code 400 (Bad Request).

\subsection{Configuration Parameters}\label{sec:config}
We augment the deployment syntax of \Jolie{} to support \emph{configuration
parameters} for our \lstcode{http} protocol. Specifically, these can be
accessed through \emph{(assign)} and \emph{(alias)} statements that can be written
inside a code block immediately after declaring the \lstcode{http} protocol in a port.
For instance, consider the following input port definition:
\begin{lstlisting}
inputPort MyInput {
  /* ... */
  Protocol: http {
    .default = "d"; .debug = true;
    .method -> m
  }
}
\end{lstlisting}
The code above would set the \lstcode{default} parameter to \lstcode{"d"}, set the
\lstcode{debug} parameter to \lstcode{true}, and bind the \lstcode{method}
parameter to the value of variable \lstcode{m} in the current \Jolie{}
process instance.

We briefly describe some configuration parameters.
All of them can be modified at runtime using the standard \Jolie{} constructs
for dynamic port binding, from~\cite{M10}, which we omit 
here.
Parameter \lstcode{default} allows to mark an operation as a special
fallback operation for receiving messages that cannot be handled by any
other operation defined in the interface of the enclosing input port.
Parameter \lstcode{cookies} allows to store and retrieve data from browser 
cookies, by
mapping cookie values in HTTP messages to subnodes in \Jolie{} messages.
Parameter \lstcode{method} allows to read/write the $Method$ field of HTTP messages. Parameter \lstcode{statusCode} gives read/write access to HTTP status codes (default status codes apply depending on the request method, e.g., the successful response to a GET request containing data has status code 200).
Parameter \lstcode{format} can be used as a default setting for the data format to use as previously discussed (e.g., \lstcode{json}, \lstcode{xml}).
The parameter \lstcode{alias} allows to map values inside a \Jolie{} message to 
resource paths in the HTTP message, to support interactions with REST services.
Parameter \lstcode{redirect} gives access to the \lstcode{Location} header in 
HTTP,
allowing to redirect clients to other locations.
The parameter \lstcode{cacheControl} allows to send directives to the client on 
how the
responses sent to it should be cached.
Finally, parameter \lstcode{debug} allows to print the HTTP messages sent and received
through the network on screen.

\subsection{Examples}\label{sec:http_examples}
We report some examples about how our \lstcode{http} protocol
implementation integrates with some standard mechanisms of web technologies.

\begin{example}[Access from web browsers]
\label{example:sum_http}
Let us consider a modification of the sum service from Example~\ref{example:sum}, where
we change the input port to use the \lstcode{http} protocol that we developed:
\begin{lstlisting}
type SumT:void { .x:int .y:int }

interface SumIface { RequestResponse: sum(SumT)(int) }

inputPort SumInput {
Location: "socket://localhost:8000"
Protocol: http
Interfaces: SumIface
}

main
{
  sum( req )( resp ) {
    resp = req.x + req.y
  }
}
\end{lstlisting}
Now, our implementation of \lstcode{http} allows us to access
the service above in multiple ways.
The most obvious is to write a Jolie client using an output port with the \lstcode{http} protocol.
A more interesting way is to use a web browser.
For example, we can use the service by passing parameters through a querystring;
navigating to the following URL is valid:
\begin{center}
\url{http://localhost:8000/sum?x=2&y=3}
\end{center}
Accessing the URL above would show the following content on the browser:
\setXML
\begin{lstlisting}
<sumResponse>5</sumResponse>
\end{lstlisting}
\setJolie
The standard format used for responses is XML, as above.
Responses from Jolie services using \lstcode{http} can of course also be themed
using, e.g., HTML and Javascript (we refer to the online documentation for more
information about this aspect~\cite{jolie:website}).

Another possibility is to use HTML forms, such as the one that follows:
\setHtml
\begin{lstlisting}
<form action="sum" method="GET">
  <input type="text" name="x"/>
  <input type="text" name="y"/>
  <input type="submit"/>
</form>
\end{lstlisting}
\setJolie
The content displayed as a response in the web browser would be the same XML
document as before.

We also offer support for AJAX programming. The following Javascript snippet
calls the \lstcode{sum} operation using jQuery~\cite{jquery}: first, it reads
the values for \lstcode{x} and \lstcode{y} from two text fields (respectively identified in the DOM by
the names \lstcode{x} and \lstcode{y}); then, it sends their values
to the Jolie service by encoding them as a JSON structure; and, finally, it displays
the response from the server in the DOM element with id \lstcode{result}:
\setJavascript
\begin{lstlisting}
$.ajax(
  'sum', { x: $('#x').val(), y: $('#y').val() },
  function( response ) { $("#result").html( response ); }
);
\end{lstlisting}
\setJolie

Our implementation of the \lstcode{http} protocol for Jolie auto-detects the format of
messages sent by clients, so the sum service does not need to distinguish among all the different
access methods shown above: they are all handled using the same Jolie code.
\end{example}

\begin{example}[Accessing REST services]
We exemplify how to access REST services, where resources are identified by URLs, using
our configuration parameters.
In this example we invoke the DBLP server, which provides bibliographic information
on computer science articles~\cite{dblp}.
We use DBLP to retrieve the BibTeX entry of an article, given the dblp key of
the latter (i.e., the identifier of such article in dblp).
The code follows:
\begin{lstlisting}
include "console.iol"

type FetchBib:void { .dblpKey:string }

interface DBLPIface {
RequestResponse: fetchBib( FetchBib )( string )
}

outputPort DBLP {
Location: "socket://dblp.uni-trier.de:80/"
Protocol: http { 
  .osc.fetchBib.alias = "rec/bib2/%!{dblpKey}.bib";
  .format = "html" }
Interfaces: DBLPIface
}

main
{
  r.dblpKey = args[0];
  fetchBib@DBLP( r )( bibtex );
  println@Console( bibtex )()
}
\end{lstlisting}
In the example above, we start by importing the \lstcode{Console} service from
the Jolie standard library.
We then declare an output port towards the DBLP server.
The interesting part here is the usage of parameter
\lstcode{osc.fetchBib.alias}, which passes to our implementation
a configuration for parameter \lstcode{alias} for operation
\lstcode{fetchBib} (\lstcode{osc} stands for operation-specific configuration and is used for configuration parameters that make sense when associated to an operation).
The value of the alias for operation \lstcode{fetchBib} specifies how to map
calls for that operation to resource paths that the DBLP server understands.
The interface offered by DBLP for retrieving bibtex entries is REST-based,
with paths rooted at ``rec/bib2/''.
As an example, assume that we wanted to retrieve the bibtex entry for the book ``The C Programming Language'' by Kernighan and Ritchie~\cite{KR78}. Its dblp
key is ``books/ph/KernighanR78''; therefore, the bibtex entry can be accessed at the URL:
\begin{center}
http://dblp.uni-trier.de/rec/bib2/books/ph/KernighanR78.bib
\end{center}
In our implementation, we capture this kind of patterns for REST paths by providing
a syntax for replacing parts of paths with the value of a subnode in a request message, based on URI templates.
For instance, the term \lstcode{\%!\{dblpKey\}} in the
alias for operation \lstcode{fetchBib}
means that that part of the path will be replaced with value of the sub node \lstcode{dblpKey}
in messages sent for that operation on port \lstcode{DBLP}.
The behaviour of the service is simple: we invoke operation \lstcode{fetchBib} reading the dblp key we
want from the first command line argument that Jolie is invoked with; then, we print the received
bibtex entry on screen.

An extended version of this example is deployed as a tool at~\cite{fm:dblp}.
\end{example}

\begin{example}[Parameter mobility]
\label{ex:param_mobility}
Configuration parameters for ports can be dynamically transmitted and used for binding services at runtime.
For example, our DBLP client may invoke a service registry to get the correct location and parameters to invoke the DBLP service as follows:
\begin{lstlisting}
/* Interface information is as before */

outputPort DBLP {
Interfaces: DBLPIface
}

outputPort Registry { /* ... */ }

main
{
  getBinding@Registry( "DBLP" )( DBLP );
  r.dblpKey = args[0];
  fetchBib@DBLP( r )( bibtex );
  println@Console( bibtex )()
}
\end{lstlisting}
In this example, the DBLP client starts by invoke an external \lstcode{Registry} service on operation \lstcode{getBinding} to dynamically discover how the DBLP service can be contacted. The registry is programmed to return the same data structure that was previously statically defined inside of the client. We refer the reader to~\cite{MGZ14} for a more detailed description of dynamic binding; here, the point is to show that our extension to configuration parameters does not lead to any additional complexity.
\end{example}

%% file: leonardo.tex
In Example~\ref{example:sum_http} we have seen how to make operations in a Jolie
service accessible by invokers using HTTP (in the example, operation \lstcode{sum}):
any operation in a Jolie service can be exposed to HTTP clients
just by changing the protocol of its related input port(s) to \lstcode{http}.
This technique covers scenarios in which the interface that we want to expose over
HTTP is statically defined as a finite set of operations, which is the typical
situation when using service-oriented technologies such as Jolie or WS-BPEL~\cite{bpel}.
However, web servers do not fall into this category. A web server allows clients
to access files, e.g., web pages, images, and JavaScript libraries.
Since files can be created and deleted during execution, we cannot statically map
each single file to an operation name as would be required by our methodology
in \S~\ref{sec:http}. In this section, we discuss how to deal with
this kind of situations by introducing default operations; these will form the basis for our development of more structured REST-oriented routers in \S~\ref{sec:rest}.

%
Default operations bridge the mutating nature of dynamic resource sets that web servers
have to offer (such as parts of a filesystem) to the static operation names used
in processes.
Specifically, a default operation is a special
operation marked as a fallback in case a client sends a request message for an
operation that is not statically defined by the service. In this case, the message
is wrapped in the following data structure (we omit some subnodes not relevant
for this discussion):
\begin{lstlisting}
type DefaultOperationHttpRequest:void {
        .requestUri:string
        .data:undefined
}
\end{lstlisting}
where \lstcode{requestUri} is the URI of the resource that has been requested
by the client and \lstcode{data} is the data content of the message.

A default operation is set through the parameter \lstcode{default} of the \lstcode{http} protocol,
and can either be associated to the $Method$ field of incoming HTTP messages or be defined 
as a ``catch-all'' operation in case no other more specific operation can be found (in the latter case, the method used in the message can then be retrieved as a variable).
For example, the following configuration states that requests for undefined 
operations with HTTP method \lstcode{PUT} 
should be handled by operation \lstcode{put},
requests for undefined operations with method \lstcode{GET} should be handled by operation 
\lstcode{get}, and
all other requests for undefined operations should be handled by operation \lstcode{d}.
\begin{lstlisting}
Protocol: http {
  .default = "d";
  .default.get = "get";
  .default.put = "put"
}
\end{lstlisting}

\begin{example}[Leonardo Web Server]
Parameter \lstcode{default} allows us to easily model a simple web server:
whenever we receive a request for the default operation, we try to find a
file in the local filesystem that has the same name as the operation
originally requested by the client.
We have used this mechanism to implement Leonardo~\cite{leonardo:website},
a web server implementation written in pure \Jolie{}.
For clarity, here we report a simplified version. The
entire implementation of Leonardo consists of only about 80 LOCs, showing that
our language pushes many of the details of dealing with HTTP to the underlying implementation;
many of these details can be accessed through configuration parameters when needed.
Leonardo can be downloaded at~\cite{leonardo:website}.
\begin{lstlisting}
/* ... */

interface MyInterface {
RequestResponse:
  d( DefaultOperationHttpRequest )( undefined )
}

inputPort HTTPInput {
Location: "socket://localhost:80/"
Protocol: http { .default = "d" /* ... */ }
Interfaces: MyInterface
}

main {
  d( req )( resp ) {
    /* ... */
    readFile@File( req.requestUri )( resp )
  }
}
\end{lstlisting}
Above, we have set the \lstcode{default} parameter for the \lstcode{http}
protocol in input port \lstcode{HTTPInput} to operation \lstcode{d}.
Therefore, when a message for an unhandled operation is received through input
port \lstcode{HTTPInput}, it will be managed by the implementation of operation
\lstcode{d}. The body of the latter invokes operation
\lstcode{readFile} of the \lstcode{File} service from the \Jolie{} standard
library, which reads the file with the same name as the originally
requested resource (\lstcode{req.requestUri}). Finally, the data read from the
file (\lstcode{resp}) is returned back to the client.
\end{example}

\begin{example}[CRUD Web Servers]
\label{example:crud}
We extend Leonardo to a simple web server supporting CRUD operations (Create, Read, Update, Delete).
As usual, we map create and update to \lstcode{PUT} requests, read to \lstcode{GET}, and delete to
\lstcode{DELETE}.
The code follows:
\begin{lstlisting}
/* ... */

interface MyInterface {
RequestResponse:
  get( DefaultOperationHttpRequest )( undefined )
  put( DefaultOperationHttpRequest )( void )
  delete( DefaultOperationHttpRequest )( bool )
}

inputPort HTTPInput {
Location: "socket://localhost:80/"
Protocol: http {
  .default.get = "get";
  .default.put = "put";
  .default.delete = "delete"
}
Interfaces: MyInterface
}

main {
  [ get( req )( resp ) {
    readFile@File( req.requestUri )( resp )
  } ] { nullProcess }
  
  [ put( req )() {
    f.filename = req.requestUri;
    f.content -> req.data;
    writeFile@File( f )( resp )
  } ] { nullProcess }
  
  [ delete( req )( resp ) {
    delete@File( req.requestUri )( resp )
  } ] { nullProcess }
}
\end{lstlisting}
In the code above, \lstcode{GET} requests are served by operation \lstcode{get}, which reads the
requested file
and replies with its content. Similarly, operation \lstcode{put} uses the Jolie standard library to write a file
with the data sent by the invoker, and operation \lstcode{delete} deletes a file from the 
filesystem.
\end{example}

%% file: sessions.tex
A main aspect of web-based information systems is the modelling of \emph{sessions},
which allow to relate different incoming messages to the same logical ``conversation''.
In this section, we present how to program sessions over HTTP with our extension of the Jolie 
language. A major benefit is that sessions are process-aware: the order in which
messages are sent and received over different operations is syntactically explicit,
and it is enforced without requiring bookkeeping variables.

\subsection{Binary sessions}
\label{sec:binary}
We start by addressing binary sessions, i.e., sessions with exactly two
participants~\cite{HVK98}.
Consider the scenario mentioned in the Introduction about a Research Information Service (RIS),
where the RIS allows users to
add a publication to a repository after having successfully logged in.
This structure is expressed by the following behaviour:
\setJolie
\begin{lstlisting}[label={lst:binary_behav}]
login( cred )( r ) { checkCredentials };
addPub( pub )
\end{lstlisting}
Above, \lstcode{login} is a request-response operation that, when invoked, checks the received credentials
by calling the subprocedure \lstcode{checkCredentials}. If the latter does not throw a fault,
the process proceeds by making operation \lstcode{addPub} available.

Suppose now that, e.g., two users are logged in at the same time in a service with the behaviour above.
The service would have then two separate process instances, respectively dedicated to handle the two
clients. When a message for operation \lstcode{addPub} arrives in this situation,
how can we know if it is from the first user or the second?
We address this kind of issues by using correlation sets, as defined in~\cite{MC11}.
A correlation set declares special variables that identify an internal service
process from the others.
In our example we use the following correlation set declaration:
\begin{lstlisting}
cset { userKey: addPub.userKey }
\end{lstlisting}
Above, we used the \lstcode{cset} keyword to declare a correlation set consisting of variable \lstcode{userKey}.
We will use \lstcode{userKey} to distinguish users that have logged in.
Variable \lstcode{userKey} is associated to the subnode \lstcode{userKey} in incoming messages for operation
\lstcode{addPub}. This means that whenever a message for operation \lstcode{addPub} is received from the network,
Jolie will assign the message to the internal running process with the same value for the correlation variable
\lstcode{userKey}.
We can now write a working implementation of the service:
\begin{lstlisting}
inputPort RISInput {
/* ... */
Protocol: http
}

cset { userKey: addPub.userKey }

define checkCredentials { /* ... */ }
define updateDB { /* ... */ }

main
{
	login( cred )( r ) {
		checkCredentials;
		r.userKey = csets.userKey = new
	};
	addPub( pub );
	updateDB
}
\end{lstlisting}
Our RIS allows the creation of new processes by invoking operation \lstcode{login}.
If the procedure \lstcode{checkCredentials} does not throw a fault,
then the process creates a
fresh value for the correlation variable \lstcode{csets.userKey} using the \lstcode{new}
keyword. The process sends the value of \lstcode{csets.userKey} back to the client through variable \lstcode{r}.
Then, the process waits for an invocation of operation \lstcode{addPub} and stores the incoming
message in variable \lstcode{pub}.
The correlation set declaration in the program
guarantees that only invocations with the same user key as that returned by operation
\lstcode{login} will be given to this process.
We finally update the internal database of the RIS using the (unspecified) procedure \lstcode{updateDB}.

\subsubsection{Integrating cookies with correlation sets}
Our implementation of the RIS requires clients to write the \lstcode{userKey} as a subnode in the 
messages they send to operation \lstcode{addPub}.
Since this may be cumbersome in the case of many operations that require correlation, web applications
typically use HTTP cookies to store this kind of information.
Our \lstcode{http} protocol integrates cookies with message correlation through the \lstcode{cookies}
parameter, which allows to map cookies to subnodes in Jolie variables.
We change the definition of input port \lstcode{RISInput} to the following:
\begin{lstlisting}
inputPort RISInput {
/* ... */
Protocol: http { .cookies.userKeyCookie = "userKey" }
}
\end{lstlisting}
The parameter assignment \lstcode{.cookies.userKeyCookie = "userKey"} instructs
our \lstcode{http} protocol implementation to store the value of the cookie
\lstcode{userKeyCookie} in subnode \lstcode{userKey} for incoming messages, and vice versa for outgoing
messages.

In general, our \lstcode{http} extension
allows developers to abstract from where correlation data is encoded when programming a service behaviour.
For instance, an important disadvantage of using cookies to store correlation data is that this breaks the statelessness constraint of REST interactions~\cite{F00}.
Instead of using a cookie, the web user interface may also send
the value for a correlation variable in other ways, e.g., embedded a hyperlink, a JSON or XML subnode, or an element in an HTML form 
encoding.
Our extension transparently support these different methods without requiring
changes in the behaviour of a service. We discuss the usage of hyperlinks to keep track of process execution in \S~\ref{sec:rest}.

\subsection{Multiparty Sessions}\label{sec:multiparty}
As far as binary sessions are concerned, there is not much difference between standard session
identifiers as used, e.g., in PHP, and correlation sets, aside from the fact that the generation
and sending of correlation variables is explicit programmed in Jolie behaviours.
However, correlation sets are more expressive when it comes to providing
(i) compound session identifier based on multiple values, as in BPEL~\cite{bpel}, and
(ii) multiple identifiers for the same process.
We are particularly interested in the second aspect, since it allows us to model \emph{multiparty} sessions,
i.e., sessions with more than two participants.

Multiparty sessions are useful when considering scenarios with multiple actors that need to be coordinated
to reach a common goal.
As an example, we extend our RIS implementation to deal with a use case from the Pure software by
Elsevier~\cite{pure}.
In Pure, when a user (e.g., a research scientist) adds a publication, a moderator (e.g., the head
of the scientist's department) has to be notified of the change. Then, the moderator has to choose
whether to approve or reject the newly added publication for confirmation in the database, after
reviewing the data inserted by the user.
We show the code for this multiparty version of our RIS implementation in the following:
\begin{lstlisting}
inputPort RISInput {
/* ... */
Protocol: http { .cookies.userKeyCookie = "userKey" }
}

outputPort Logger { /* ... */ }
outputPort Moderator { /* ... */ }

cset { userKey: addPub.userKey }
cset { modKey: approve.modKey reject.modKey }

define checkCredentials { /* ... */ }
define updateDB { /* ... */ }

main
{
	login( cred )( r ) {
		checkCredentials;
		r.userKey = csets.userKey = new
	};
	addPub( pub );
	noti.bibtex = pub.bibtex;
	noti.modKey = csets.modKey = new;
	{ log@Logger( pub.bibtex ) | notify@Moderator( noti ) };
	[ approve() ] {
		log@Logger( "Accepted " + pub.bibtex );
		updateDB
	}
	[ reject() ] {
		log@Logger( "Rejected " + pub.bibtex )
	}
}
\end{lstlisting}
Above, we have added the output ports \lstcode{Logger}, an external service
that maintains a log of actions that we assume the user can read, and \lstcode{Moderator},
an external service playing the role of the moderator in our scenario.
We have also added a new correlation set for variable \lstcode{modKey} (moderator key), which
we use to track incoming messages from the moderator of a session. The correlation set declares
also that the moderator may use \lstcode{modKey} to invoke operations \lstcode{approve} and
\lstcode{reject}.
In the behaviour, the code is unchanged until after we receive an invocation for operation \lstcode{addPub}.
Now, after we receive a request for operation \lstcode{addPub}, we prepare a notification \lstcode{noti}
for the moderator containing (i) the descriptor of the
publication (we assume that it is given by the user in BibTeX format),
and (ii) the moderation key \lstcode{modKey} (which is instantiated as a fresh value with the 
keyword
\lstcode{new}).
Then we use the parallel construct of Jolie to concurrently
send a message to, respectively, the \lstcode{Logger} on operation
\lstcode{log} (to log the user's request) and the \lstcode{Moderator} on operation \lstcode{notify} (to
notify the moderator of the user's request).
The process now enters into an input choice on operations \lstcode{approve} and
\lstcode{reject}, which can be invoked only by the moderator; this is because
the correlation set declaration of variable \lstcode{modKey} requires it to be present
for invocations of these operations, and we sent the value of \lstcode{modKey} only to the moderator.
If \lstcode{approve} is invoked, then we log the approval and we update the database of publications.
Otherwise, if \lstcode{reject} is invoked, we log the rejection only.

%% file: architectural.tex
In the previous sections, we focused separately on how to use our extension of Jolie
to program web servers (\S~\ref{sec:leonardo}) and structured process-aware sessions
(\S~\ref{sec:sessions}).
Typically, a real-world web architecture has to deal with both aspects. In this section,
we show how they can be combined in our context by building multi-layered architectures.

\subsection{Aggregation}
\label{sec:aggregation}
A simple way of designing a service that serves content \emph{and} provides
process-aware sessions is to combine the respective operations in the same behaviour as
an input choice.
Consider the following code:
\begin{lstlisting}
/* ... */
main
{
  [ get( req )( resp ) { /* ... */ } ] { nullProcess }
  [ login( cred )( r ) { /* ... */ } ] { /* ... */ }
}
\end{lstlisting}
Above, we assume that the ports and correlation sets are configured by merging
the configurations found in Example~\ref{example:crud}
and the RIS implementation in \S~\ref{sec:multiparty}.
Then, operation \lstcode{get} would serve HTML and JavaScript files to clients,
which could also invoke operation \lstcode{login} to access the behaviour of the RIS.

While combining the code for a web server with that of sessions with complex structures
as done above is simple, in the long term it also leads to code that
is hard to maintain due to poor separation of concerns: all concerns are mixed in the same
service.
Ideally, separate concerns should be addressed by separate services.
This methodology, however, raises the question of how services addressing separate concerns
can be composed together as a single system that clients can access without knowing the
inner complexity of the system.
We tackle this issue by integrating our \lstcode{http} protocol implementation with
the notion of service aggregation found in Jolie~\cite{M10,MGZ14}.

Aggregation is a Jolie primitive that allows a service
to expose the interfaces of other services on one of its input ports, in addition to its own interfaces.
In the remainder, we refer to the service using aggregation as aggregator and to the
services it aggregates as aggregated services.
The semantics of aggregation is a simple generalisation of the mechanism
used in proxy services:
when a message from the network reaches an aggregator, the aggregator
checks whether the message is for an operation in (i) one of its own interfaces
or (ii) the interfaces of an aggregated service. In the first case,
the message is given to the behaviour of the aggregator; in the second case,
the aggregator forwards the message to the aggregated service providing the
operation requested in the message.

\begin{figure}[t]
\begin{center}
\includegraphics[scale=0.5]{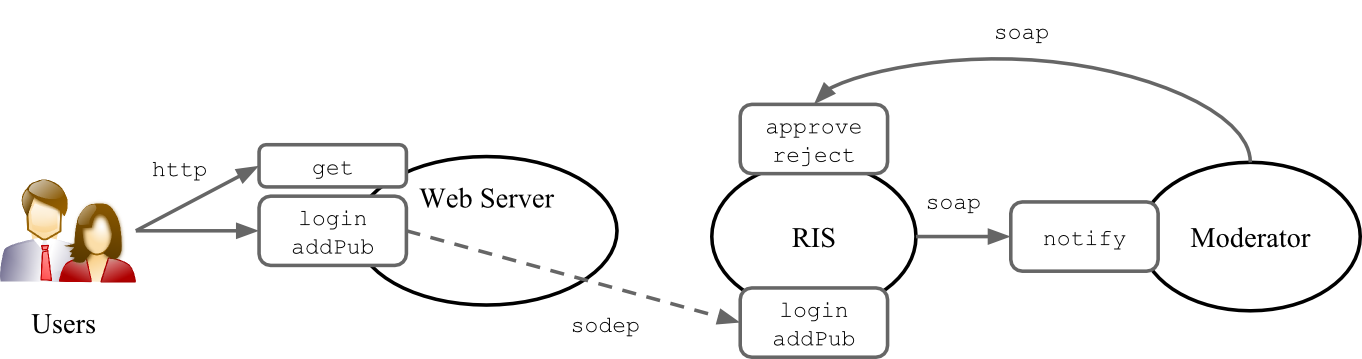}
\end{center}
\caption{Architecture of the RIS scenario.}
\label{fig:ris}
\end{figure}

Using aggregation in combination with our \lstcode{http} protocol
we can easily build a multi-layered web architecture for our RIS scenario, where services
communicate using different protocols as needed.
We depict the architecture in Figure~\ref{fig:ris}, where circles represent services,
rectangles represent the interfaces exposed by services, full arrows represent dependencies
from actors (users or services) to services, and dashed arrows represent aggregations;
each arrow is annotated with the protocol used for communications.
We comment the architecture.
Users can access the web server using a web browser, through the \lstcode{http} protocol.
Requests for files, intended to be for the user interface (e.g., HTML pages),
are handled directly by the web server through operation \lstcode{get}.
Instead, invocations of operations \lstcode{login} and \lstcode{addPub} are forwarded to
the RIS by aggregation.
The web server and the RIS communicate using the \lstcode{sodep} protocol,
for performance (\lstcode{sodep} is a binary protocol).
As in \S~\ref{sec:multiparty}, the RIS uses an additional service, Moderator, to decide
whether publications should be accepted into the system. The RIS and the Moderator services
communicate using the \lstcode{soap} protocol.
Below, we exemplify how our architecture can be implemented. We assume that the Moderator service
is externally provided, and focus instead on the web server and the RIS.

\paragraph{Web server}
The code of the web server follows:
\begin{lstlisting}
/* ... */

outputPort RIS {
Location: "socket://www.ris-example.com:8090/"
Protocol: sodep
Interfaces: RISIface
}

inputPort WebServerInput {
Location: "socket://www.webserver-example.com:80/"
Protocol: http {
  .default.get = "get";
  .cookies.userKeyCookie = "userKey"
  /* ... */
}
Interfaces: GetIface
Aggregates: RIS
}

main {
  get( req )( resp ) {
    /* ... */
    readFile@File( req.requestUri )( resp )
  }
}
\end{lstlisting}
Our web server implements only the operation \lstcode{get}, which serves static files to clients.
It also aggregates the RIS by using the aggregation instruction \lstcode{Aggregates: RIS} in its
input port, where \lstcode{RIS} is an output port pointing to the RIS.
Therefore, all invocations from users for the operations offered by the RIS will be automatically
forwarded to the latter.
Observe that output port
\lstcode{RIS} uses the \lstcode{sodep} protocol: our implementation automatically takes care
of translating incoming HTTP messages from users destined to the RIS into binary
\lstcode{sodep} messages. In general, the programmer does not need to worry about data format
transformations in our extension of the Jolie language: messages are implicitly converted
to/from the HTTP format as needed.

\begin{remark}
Here, we showed the code for the web server modified to aggregate the RIS for clarity purposes. In real-world production environments, the practice of rewriting the web server component every time comes with the unnecessary risk of introducing bugs. Therefore, in such environments, the web server is deployed as an autonomous service, such that important updates to it from the Leonardo project (or any other web server project based on Jolie) can be immediately applied. In this kind of set-up, the configuration information on which services should be aggregated by Leonardo is kept in a separate configuration file.
\end{remark}

\paragraph{RIS (Research Information Service)}
The code for the RIS is the same as that shown in \S~\ref{sec:multiparty}, with the exception
that we now use \lstcode{sodep} as communication protocol and that we removed
the usage of the external service \lstcode{Logger} for simplicity:
\begin{lstlisting}
inputPort RISInput {
Location: "socket://www.ris-example.com:8090/"
Protocol: sodep
Interfaces: RISIface
}

outputPort Moderator {
Location: "socket://www.moderator-example.com:8080/"
Protocol: soap
Interfaces: ModeratorIface
}

cset { userKey: addPub.userKey }
cset { modKey: approve.modKey reject.modKey }

define checkCredentials { /* ... */ }
define updateDB { /* ... */ }

main
{
	login( cred )( r ) {
		checkCredentials;
		r.userKey = csets.userKey = new
	};
	addPub( pub );
	noti.bibtex = pub.bibtex;
	noti.modKey = csets.modKey = new;
	notify@Moderator( noti );
	[ approve() ] {
		updateDB
	}
	[ reject() ] {
		/* ... */
	}
}
\end{lstlisting}

\subsection{Evolvability}
\label{sec:evolvability}
Implementing multi-layered web architectures using our approach, i.e., combining
\lstcode{http} with aggregation, results in systems that are robust wrt future
modifications, or their \emph{evolution}.
We distinguish between \emph{vertical} and \emph{horizontal modifications}, which respectively
represent modifications that influence an existing chain of aggregations or new ones.

A vertical modification is a modification of an interface aggregated by another service.
In our example, changing the code of the RIS to add, remove, or change the type of an operation
in interface \lstcode{RISIface} would be a
vertical modification, because \lstcode{RISIface} is aggregated by the web server.
Vertical modifications do not require any intervention on the rest of the architecture,
as aggregation is a parametric mechanism: the web server simply needs to be restarted to read the new
definition of interface \lstcode{RISIface}.

\begin{figure}
\begin{center}
\includegraphics[scale=0.5]{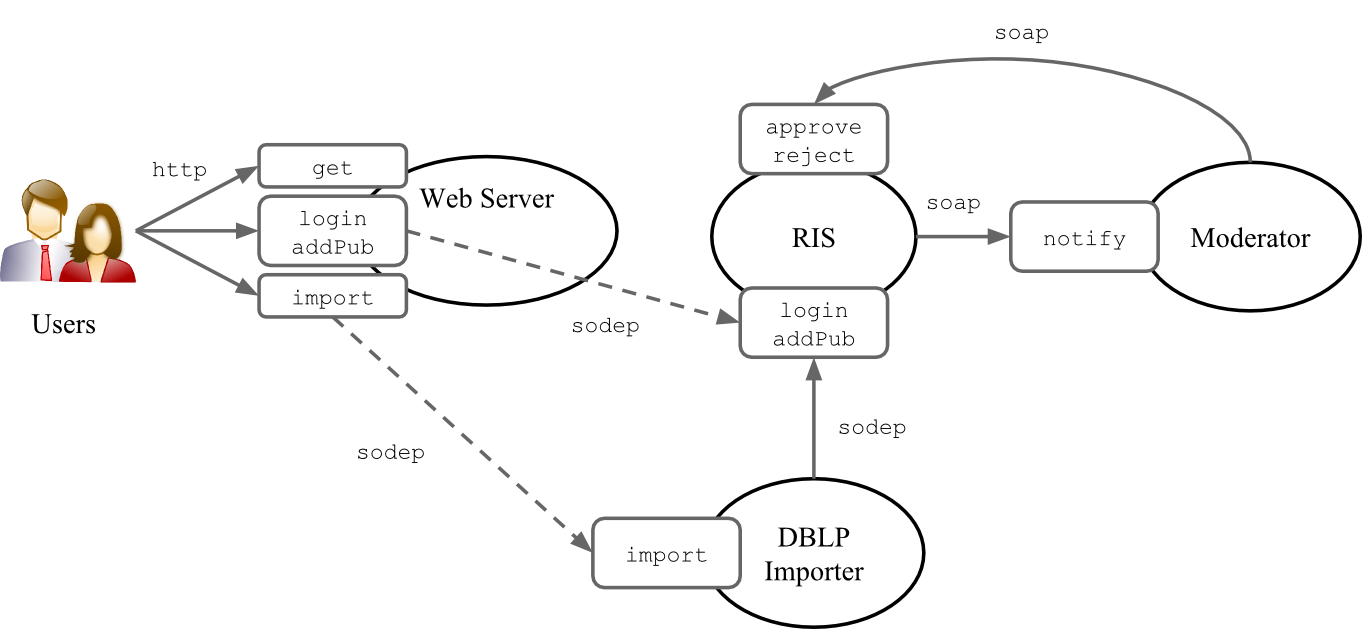}
\end{center}
\caption{Architecture of the RIS scenario with DBLP importer.}
\label{fig:ris_import}
\end{figure}

Horizontal modifications deal with the addition or removal of operations without requiring
an intervention on the behaviour of existing services.
Assume that, as an example, we wanted to add the possibility to import publications from
the DBLP bibliography service to our RIS by offering a new operation called \lstcode{import}.
We could implement this new feature by changing the code of the RIS, both its interface and behaviour.
However, in some scenarios this may not be possible, e.g., the RIS may be a black box to which we do not
have access (third-party proprietary code), or the RIS cannot be modified due to quality or security
regulations.
We deal with this kind of situations by developing the new operations we need in a new service,
and then by aggregating this service together with the RIS in the web server.
The resulting situation in our example scenario is depicted in Figure~\ref{fig:ris_import}.
The only difference between Figure~\ref{fig:ris_import} and our previous
architecture from Figure~\ref{fig:ris} is the presence of a new service, called Importer,
offering operation \lstcode{import}; the web server now aggregates Importer together with the RIS, to
make operation \lstcode{import} accessible by users through web browsers.
We report the updated code for the web server and the importer.

\paragraph{Web server}
For the web server, we simply need to add an output port towards the importer service and
aggregate it in the input port of the server. We report only the code interested by our changes, the rest
remains the same as in \S~\ref{sec:aggregation}.
\begin{lstlisting}
/* ... */

outputPort Importer {
Location: "socket://localhost:8009/"
Protocol: sodep
Interfaces: ImporterIface
}

outputPort RIS { /* ... */ }

inputPort WebServerInput {
/* ... */
Aggregates: RIS, Importer
}

/* ... */
\end{lstlisting}
By changing the web server as done above, invocations for operation
\lstcode{import} will now be redirected to the importer service.

Observe that, by using aggregation, all the invocations from the web client to the aggregated
services pass through the web server. This implies that our programming methodology
respects the standard Same Origin Policy by design, allowing the web application run by users 
to access the aggregated services regardless of where the latter are located.

\paragraph{Importer service}
The code for the importer service follows:
\begin{lstlisting}
/* ... */

outputPort DBLP {
Location: "socket://dblp.uni-trier.de:80/"
Protocol: http {
	.osc.fetchBib.alias = "rec/bib2/%!{dblpKey}.bib"
	/* ... */
}
Interfaces: DBLPIface
}

outputPort RIS { /* ... */ }

inputPort ImporterInput {
Location: "socket://localhost:8009/"
Protocol: sodep
Interfaces: ImporterIface
}

main
{
	import( request );
	
	dblpReq.dblpKey = request.dblpKey;
	fetchBib@DBLP( dblpReq )( result );
		
	addReq.bibtex = result;
	addReq.userKey = request.userKey;

	addPub@RIS( addReq )
}
\end{lstlisting}
The importer service offers a single operation, \lstcode{import}, which
takes as input a message containing two subnodes:
\lstcode{dblpKey}, the dblp key of the publication to import from DBLP,
and \lstcode{userKey}, which must be a valid user key for a session inside of the RIS.
The idea is that a user has to invoke operation \lstcode{login} before using operation
\lstcode{import}, thus opening a session in the RIS,
and that she then invokes \lstcode{import} with the \lstcode{userKey} it got
as a response from \lstcode{login}.
After receiving a message for operation \lstcode{import}, the importer service proceeds
by invoking the DBLP service to retrieve the BibTeX record stored therein for the dblp key passed
by the user. Finally, after retrieving the BibTeX record, the importer asks the RIS to add it
through operation \lstcode{addPub}.

%% file: rest.tex
So far, we have considered architectures based on the style of Web Services~\cite{A04,wsi}, where our \lstcode{http} extension is used to bind HTTP messages to verb-based services.
In this section, we explore how to use the Jolie-HTTP binding to follow the REpresentational State 
Transfer architectural style (REST)~\cite{F00}.
%

\subsection{REpresentational State Transfer (REST)}
The REST architectural style is a collection of principles for the design and development of web systems,
based on the key abstract concept of \emph{resource}~\cite{F00}.
These principles are aimed at promoting the 
scalability and reusability of services by constraining how they should interact. Such constraints reduce coupling 
and allow for the use of intermediary services to improve, among other aspects, performance, security, and 
integration.
We briefly present the principles of the REST style in the following (see~\cite{P14} for a more comprehensive overview of REST and its 
adoption).
\begin{enumerate}
\item \emph{Resource Identification.} Interactions with a RESTful service happen by referring to the resources that it 
exposes. Resources are globally identified by URIs~\cite{uri}.
\item \emph{Uniform Interface.} The operations that can be used on resources are fixed.
For example, the uniform interface offered by HTTP includes the operations (called methods) GET, POST, 
PUT, and DELETE. GET retrieves a representation of the state of a resource. 
POST submits some data to modify the state of a resource.
PUT creates a new resource. Dually, DELETE is used for deleting a resource.
The idea is that the uniform interface should be a small set of operations, yet generic enough to implement all the 
desired functionalities in a web system.
\item \emph{Self-descriptive Messages.} Each message must contain all the necessary data and metadata for the message 
to be processed. For example, metadata can be used to indicate the format used to encode the message payload, informing 
the receiver of how it should be unmarshalled.
\item \emph{Hyperlinks as the engine of application state.}
State transitions in a RESTful service are explicitly communicated.
After the service processes a request that causes a state transition, it sends hyperlinks in the response 
to inform the client of what resources it can now use.
\end{enumerate}

\subsection{Routing via URI templates}
We show how to program a RESTful service in our framework by implementing a service for online polls, inspired by 
the reference example used in~\cite{P14} to introduce the REST style.
This poll service allows clients to manage polls, where each poll has a number of options that can be chosen from. 
Technically, this is achieved by offering the resources reported in Table~\ref{table:poll_res}, where we also describe
the semantics of each HTTP method for a resource.
\input{figures/poll_res}

A naive but immediate solution for obtaining a RESTful service is to write a Jolie service that implements the Uniform 
Interface (here we limit our discussion to the operations GET, POST, PUT, and DELETE).
We call this interface \lstcode{WebIface}:
\begin{lstlisting}
interface WebIface {
RequestResponse:
get(WebRequest)(undefined), post(WebRequest)(undefined),
put(WebRequest)(undefined), delete(WebRequest)(undefined)
}
\end{lstlisting}
Each operation receives a message of type \lstcode{WebReq}, from \S~\ref{sec:leonardo}, and replies with a message of 
type \lstcode{undefined}, since the type of the responses depends on the resource that the operation is applied to.
We can implement the interface using our default operations:
\begin{lstlisting}
inputPort WebIn {
Location: "socket://www.mysrv.com:80"
Protocol: http {
	.default.get = "get"; .default.post = "post";
	.default.put = "put"; .default.delete = "delete"
}
Interfaces: WebIface
}

main {
	[ get( request )( response ) { /* ... */ } ]
	[ post( request )( response ) { /* ... */ } ]
	[ put( request )( response ) { /* ... */ } ]
	[ delete( request )( response ) { /* ... */ } ]
}
\end{lstlisting}
In the service above, each operation implements its respective HTTP method.
We now need a way of differentiating the behaviour of each operation, e.g., \lstcode{get}, depending on the resource 
requested by the client.
For this purpose, we use standard URI templates~\cite{uritemplates}. Specifically, we introduce a Jolie standard 
library service, called \lstcode{UriTemplates}, that we can use to match the resource URI sent by a client to a URI 
template and extract the parameters embedded in the resource URI (e.g., \emph{pid} in Table~\ref{table:poll_res}).
As URI templates for the poll service, we use those reported in the first column of Table~\ref{table:poll_res}. 
A naive use of service \lstcode{UriTemplates} is the following, where we exemplify how to support GET invocations for 
individual polls and votes:
\begin{lstlisting}
main {
	get( request )( response ) {
		match@UriTemplates( {
			.uri = request.requestUri,
			.template = "/poll/{pid}"
		} )( m );
		if ( m ) { // GET /poll/{pid}
			response << global.polls.(m.pid)
		} else { // GET /poll/{pid}/vote/{vid}
			match@UriTemplates( {
				.uri = request.requestUri,
				.template = "/poll/{pid}/vote/{vid}"
			} )( m );
			response << global.polls.(m.pid).votes.(m.vid)
		}
	}
}
\end{lstlisting}
As an example, in the code above, if we receive an HTTP request targeting URI \lstcode{/poll/5/vote/2}, we would get a 
positive match in Lines 10-13 and the subnodes \lstcode{pid} and \lstcode{vid} in variable \lstcode{m} would be set to 
the respective values \lstcode{5} and \lstcode{2} from the URI. In Line 14, we use these values to find the right vote 
to return to the invoker.


\subsection{The Router Service}
\label{sec:rest_router}
While functional, the approach we followed in programming the poll service suffers from poor 
separation of concerns and readability. This is because the code for determining which action we should perform, by 
matching URIs with URI templates, is mixed with the code for implementing each action.
We solve this problem architecturally, by introducing a standard router service that can be autonomously deployed and therefore modularly reused in the 
implementation of any RESTful system. The idea is to separate a RESTful service into two services: a \emph{router}, 
which deals with matching URIs to actions, and a \emph{controller}, which deals with the implementations of 
actions. (The same router may actually manage multiple controllers, but here we will focus on just one.)

\begin{figure}
\begin{center}
\includegraphics[scale=0.5]{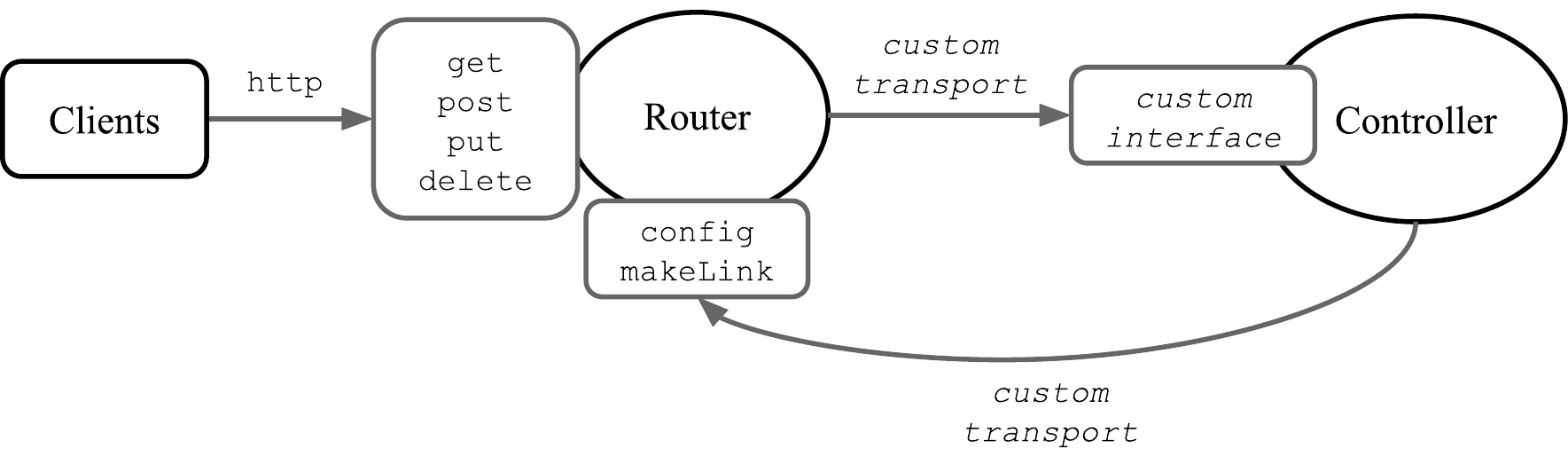}
\end{center}
\caption{Router architecture for RESTful Jolie services.}
\label{fig:router}
\end{figure}

We depict the resulting architecture for a RESTful Jolie service in Figure~\ref{fig:router}.
In the Figure, the router service is the service that exposes the uniform interface to clients. The implementation 
of actions is then delegated to the controller, which offers a custom interface defined by the programmer.
We remark two operations that the router service offers to the controller: 
\lstcode{config} and \lstcode{makeLink}. Operation \lstcode{config} is used to tell the router how client invocations 
for its uniform interface should be matched with the user-defined operations offered by the controller. Operation 
\lstcode{makeLink}, instead, is used by the controller to build hyperlinks that respect the configuration of the 
router, ensuring that clients receive hyperlinks that can be correctly routed in the future.
We impose no restriction on the transport between the router and 
the controller: it can be anything supported by Jolie, including, e.g., shared memory communications (the default, 
obtained using local locations~\cite{MGZ14}) or Web Services protocols (e.g., SOAP). The architecture can be nested, 
allowing for the composition of RESTful and/or non-RESTful services: the controller may be a router itself, or 
orchestrate operations offered by other routers.


In our poll service example, the poll service is the controller. We can use the router to rewrite the poll service as 
follows.
%
\begin{lstlisting}
init {
	config.host = "localhost:8080";
	// Resource poll
	config.resources[0] << {
		.name = "poll",
		.id = "pid",
		.template = "/poll"
	};
	// vote sub-resources of poll
	config.resources[0].resources[0] << {	 
		.name = "vote",
		.id = "vid",
		.template = "/vote"
	};
	config@Router( config )()
}

main {
	[ poll_index()( response ) {	// GET /poll
		foreach( pid : global.polls ) {
			makeLink@Router( {
				.operation = "poll_show",
				.params.pid = pid
			} )( response.href[i++] )
		}
	} ]

	[ poll_show( request )( response ) {	// GET /poll/{pid}
		findPoll;
		response.options << poll.options;
		makeLink@Router( {
			.operation = "vote_index",
			.params.pid = request.id
		} )( response.votes.href );
		for( i = 0, i < #poll.vote, i++ ) {
			response.votes.vote[i] << poll.vote[i]
		}
	} ]

	[ vote_index( request )( response ) {	// GET /poll/{pid}/vote
		findPoll;
		response.vote -> poll.vote
	} ]
	
	// etc.
}
\end{lstlisting}
In Lines 2--15, the router is configured to target the address of our controller and then to offer poll resources and their sub-resources vote. The router will then route requests to corresponding operations suffixed with special identifier to distinguish what HTTP method they should receive from. These operations are provided in the \lstcode{main} block of the controller; we put comments to indicate what kind of calls correspond to their invocations.
For example, operation \lstcode{poll\_index} is invoked by GET requests for resource /poll, which returns a list of hyperlinks towards existing polls. Observe that we build hyperlinks by invoking operation \lstcode{makeLink}, offered by the router, which allows us to abstract from the URI templates that we are using and avoid the error-prone manual writing of hyperlinks.
The rest of the service implements the behaviour described in Table~\ref{table:poll_res}.

We now move to the implementation of the router service, focusing on its most important behaviour:
\begin{lstlisting}
/* ... */

define route {
	findRoute;
	if ( !found ) {
		setErrorStatusCode
	} else {
		/* prepare invokeReq */
		invokeReq.operation = found.operation;
		invoke@Reflection( invokeReq )( response );
		statusCode = 200;
		/* set headers and status code */
	}
}

main
{
	[ get( request )( response ) {
		method = "get";
		route
	} ]

	[ post( request )( response ) {
		method = "post";
		route
	} ]

	[ makeLink( request )( response ) { /* ... */ } ]
	
	/* ... */
}
\end{lstlisting}
The router offers the Uniform Interface and the \lstcode{config} and  \lstcode{makeLink} operations (we omit code that works as in other standard frameworks, not relevant for our discussion).
The most important part is how messages are routed from the Uniform Interface to the controller. For each operation in the Uniform Interface, e.g., \lstcode{get} and \lstcode{post}, the \lstcode{route} procedure is called and looks for the appropriate route for that method in the configuration. When this is found, variable \lstcode{invokeReq} is programmed to contain the location of the controller, the data to send, and the operation to invoke (Lines 229--230). Now we need a way to invoke an operation that we do not statically know; for this, we introduce a new \lstcode{Reflection} service to the Jolie standard library that supports typical reflection operations, as in Java. In Line 231, \lstcode{Reflection} is used to invoke the operation decided at runtime on the controller. Depending on the configuration and whether the operation replies with a response or a fault, the headers and status code in the reply are set accordingly.

\subsection{RESTful Processes}\label{sec:rest_processes}

In our poll service example, the life cycle of resources is handled with a shared memory space. Hereby, we describe how a more process-oriented approach can be used by combining correlation sets with REST routing. We also report on how custom operations (without suffixes) can be used in route configurations.
The program below is a service that manages quizzes under path /quiz. The code follows:
\begin{lstlisting}
/* ... */

cset {
id:	Start.id Show.id Answer.id
		Confirm.id Giveup.id Timeout.id
}

init {
	config.routes[0] << {
		.method = "post",
		.template = "/quiz",
		.operation = "start"
	};
	config.routes[1] << {
		.method = "get",
		.template = "/quiz/{id}",
		.operation = "show"
	};
	config.routes[2] << {
		.method = "delete",
		.template = "/quiz/{id}?reason=confirm",
		.operation = "confirm"
	};
	/* other routes... */
	config@Router( config )()
}

main
{
	start( request )( response ) {	// POST /quiz
		csets.id = new;
		makeLink@Router( {
			.operation = "show",
			.params.id = "id"
		} )( response.href );
		send@Mail( /* send link to request.player */ )();
		quiz -> request.quiz
	};
	setNextTimeout@Time( TO { .message.id = csets.id } );
	provide
		[ show()( state ) {			// GET /quiz/{id}
			/* reply with quiz text and hyperlinks */
		} ]					
		[ answer( quiz.answers ) ]	// PUT /quiz/{id}/answers
	until
		[ confirm()() ]	// DELETE /quiz/{id}?reason=confirm
		[ giveup()() ]	// DELETE /quiz/{id}?reason=giveup
		[ timeout() ];	// Local memory call from Time
	
	/*
		Send e-mail with results to
		request.quizmaster and request.player
	*/
}
\end{lstlisting}
The service starts with the configuration of its own correlation mechanism and that of the router (Lines 3--26). In the \lstcode{main} procedure (the behaviour of the service), we start by offering to clients the possibility of creating a new quiz. A quiz creation request must include some text (\lstcode{request.quiz}) and the e-mail address of the player the quiz is intended for (\lstcode{request.player}). After the correlation token for the quiz is created (Line 31), we send the hyperlink to access the quiz (asked to the router in Lines 32--35) to the player (Line 36).
The process now created to handle the quiz registers a timeout (of duration \lstcode{TO}) using the \lstcode{Time} service from the Jolie standard library.
We then enter a \lstcode{provide-until} block: the operations \lstcode{show} and \lstcode{answer} are repeatedly provided to clients until either \lstcode{confirm}, \lstcode{giveup}, or \lstcode{timeout} is invoked. Each operation is commented with how it can be reached from the Web (aside from \lstcode{timeout}, which is local).
Operation \lstcode{show} is meant to be invoked by the player, who received the hyperlink to access it via e-mail when the quiz was created. Operation \lstcode{answer} is used to create (or replace) the sub-resource answers of the quiz.
When she is satisfied, the player can either \lstcode{confirm} her answers or \lstcode{giveup}, using different hyperlinks.
Finally, the quizmaster and the player are notified of the outcome.

\subsection{Integrating Javascript}
\label{sec:javascript}
The client input for manipulating or accessing resources typically has to be validated to check for errors (e.g., invalid syntax). This must be done both on the client and on the server: the former is to offer rapid feedback to the user and avoid sending wrong requests in the first place, whereas the latter is aimed at  avoiding malicious client requests. Since validation code in the client is typically written in JavaScript (unless it is trivially supported by HTML or similar declarative mechanisms), we developed an integration mechanism between Jolie and JavaScript to be able to reuse JavaScript programs in Jolie services and avoid error-prone duplication of logic. The basic idea is to enable the invocation of JavaScript programs as if they were services. We obtain this property by extending the \emph{embedding} mechanism of Jolie~\cite{MGZ14} to support the execution of JavaScript programs as sub-services of a Jolie program.

Concretely, our extension allows to bind output ports to functions offered by a JavaScript program via an \lstcode{embedded} block in the deployment part of a Jolie service. 
In our quiz example, assume that the JavaScript code used by the client for validating the quiz creation request is in a function called \lstcode{validate} that resides in a file called\lstcode{script.js}. We can bind this function to an output port as follows:
\begin{lstlisting}
interface JSIface {
  RequestResponse: validate(QuizRequest)(bool) }
outputPort JS { Interfaces: JSIface }
embedded { Javascript: "script.js" in JS }	
\end{lstlisting}
Our extension takes care of converting Jolie values to JavaScript objects and vice versa. Now, we can invoke the \lstcode{validate} function before Line 31 in our example:
\begin{lstlisting}
validate@JS( request )( ok );
if ( !ok ) { throw( MalformedRequest ) }
\end{lstlisting}

%% file: figures/poll_res.tex
\begin{table}
\centering
{\setlength\tabcolsep{0.15cm}{\setlength{\extrarowheight}{1mm}
\begin{tabular}{|l|c|c|c|c|}
\hline \multicolumn{1}{|c|}{\textbf{Path}} & \multicolumn{4}{c|}{\textbf{Actions}}\\
\hline & \multicolumn{1}{c|}{GET} & \multicolumn{1}{c|}{POST} & \multicolumn{1}{c|}{PUT} & 
\multicolumn{1}{c|}{DELETE} \\[1mm]
\hline \multirow{2}{*}{/poll} &
\multirow{2}{*}{\begin{tabular}{c}Get list \\ of polls\end{tabular}} & 
\multirow{2}{*}{Create poll} &
\multirow{2}{*}{-} & \multirow{2}{*}{-}
\\ &&&& \\[1mm]
\hline \multirow{2}{*}{/poll/\emph{\{pid\}}} &
\multirow{2}{*}{\begin{tabular}{c}Get poll \emph{pid}\end{tabular}} & 
\multirow{2}{*}{-} &
\multirow{2}{*}{
-
} &
\multirow{2}{*}{\begin{tabular}{c}Delete\\ poll \emph{pid}\end{tabular}}
\\ &&&& \\[1mm]
\hline \multirow{2}{*}{/poll/\emph{\{pid\}}/vote} &
\multirow{2}{*}{\begin{tabular}{c}Get votes\\ of poll \emph{pid}\end{tabular}} & 
\multirow{2}{*}{\begin{tabular}{c}Create vote\\ in poll \emph{pid}\end{tabular}} &
\multirow{2}{*}{-} & \multirow{2}{*}{-}
\\ &&&& \\[1mm]
\hline \multirow{2}{*}{/poll/\emph{\{pid\}}/vote/\emph{\{vid\}}} &
\multirow{2}{*}{\begin{tabular}{c}Get vote \emph{vid}\\ of poll \emph{pid}\end{tabular}} & 
\multirow{2}{*}{-} &
\multirow{2}{*}{\begin{tabular}{c}Set state\\ of vote \emph{vid}\end{tabular}} &
\multirow{2}{*}{\begin{tabular}{c}Delete\\ vote \emph{vid}\end{tabular}}
\\ &&&& \\[1mm]
\hline
\end{tabular}
}}
\caption{Resources offered by the poll service.}
\label{table:poll_res}
\end{table}

%% file: performance.tex
In this Section, we present the results of some representative performance experiments executed with our framework. The aim of these experiments is to obtain indicative information on the applicability of our work. 
All experiments were run on a server machine equipped with an i5-2500k CPU and 8GB of RAM memory.

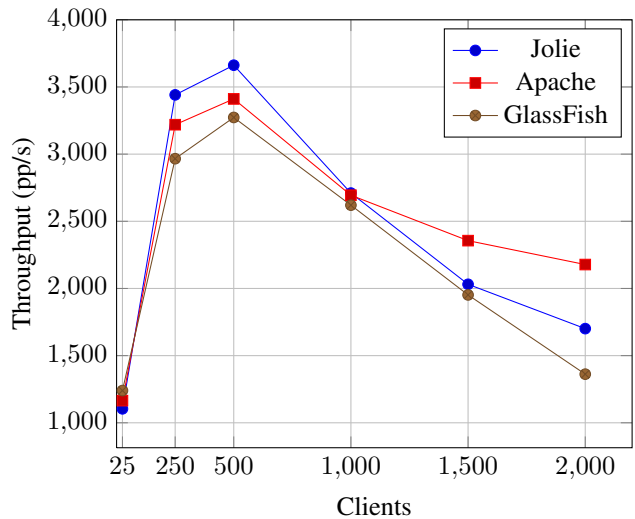
\begin{figure}
\centering
\begin{tikzpicture}
\begin{axis}[
	xlabel={Clients},
	ylabel={Throughput (pp/s)},
	legend entries={Jolie, Apache, GlassFish},
	xmin=0,
	xtick=data,
	grid=both,
	ytick={1000,1500,2000,2500,3000,3500,4000},
	ymax=4000
]
\addplot
table {results/static_jolie.dat};
\addplot table {results/static_apache.dat};
\addplot table {results/static_glassfish.dat};
\end{axis}
\end{tikzpicture}
\caption{Static page download experiment.}
\label{fig:static_experiment}
\end{figure}

\paragraph{Serving static content}
A central feature of any web system is serving static content, e.g., an HTML page, an image, or a CSS file. In this experiment, we measure the throughput (pages downloaded per second) of each framework when serving a growing number of clients a simple web page of middle size (about 1,500 bytes). The results are shown in Figure~\ref{fig:static_experiment}. In the graph, our solution is labelled Jolie. The other frameworks used in the comparison are the Apache HTTP server\footnote{\url{https://httpd.apache.org/}} and the GlassFish server\footnote{\url{https://glassfish.java.net/}}, the reference implementation of Java EE (Enterprise Edition) by Oracle.
	It is clear from the data that Jolie offers comparable performance wrt the other two frameworks.

\begin{figure}
\centering
\begin{tikzpicture}
\begin{axis}[
	xlabel={Clients},
	ylabel={Throughput (pp/s)},
	legend entries={Jolie, Apache/PHP, GlassFish/JSP},
	xtick=data,
	ymin=550,
	ymax=650,
	grid=both
]
\addplot table {results/academia_jolie.dat};
\addplot table {results/academia_apache.dat};
\addplot table {results/academia_glassfish.dat};
\end{axis}
\end{tikzpicture}
\caption{Templated page download experiment.}
\label{fig:template_experiment}
\end{figure}
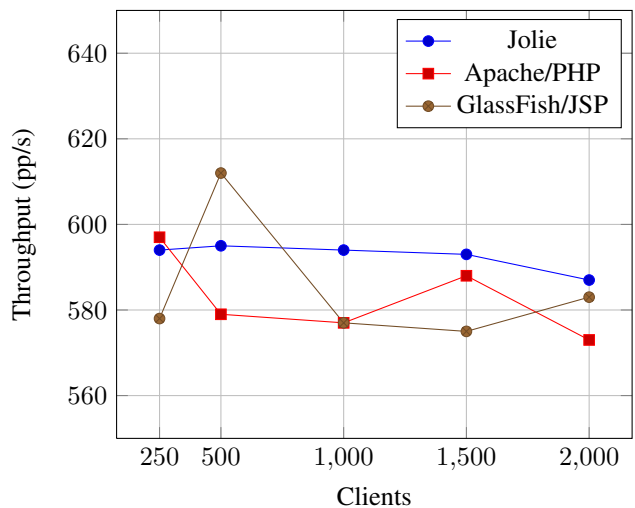

\paragraph{Templated content}
In this experiment, we measure the speed in serving dynamic pages, i.e., pages whose content is computed at runtime for each request rather than being statically stored in a file. Specifically, here the server receives the request for a page, retrieves the main page content and then sets it inside of a template containing a navigation menu and graphical layout\footnote{The template used is that of the Jolie website, with a simple header and footer: \url{http://www.jolie-lang.org/}}.
To compute the final result for the clients in Apache and GlassFish, we used PHP (version 5) and Java Server Pages (JSP) respectively.
The results, given in Figure~\ref{fig:template_experiment}, show again that the three frameworks perform with relatively close speeds. 

\paragraph{Scalability}
In our setting, replying to a client request may involve the use of many services (and hence many processes). It is therefore interesting to observe the impact that the number of services involved in processing a client request has on scalability. To this aim, we run an experiment where a client can ask a Jolie web server for a list of news; the web server then asks a collector service to retrieve (in parallel) news from a number of autonomous news services; finally, the aggregated list of news is returned to the client.
	We measure the performance of the system (as throughput) using 20, 40, 60, and 80 news services that the collector service must contact.
	The results are shown in Figure~\ref{fig:scalability_experiment}, where each curve represents the behaviour of the system when continuously receiving requests from 250, 500, 750, or 1000 clients respectively.
	As first observation, we note that the system scales well wrt the number of clients (as for the templated page experiment). When we increase the number of news services, throughput decreases more significantly. This is to be expected since each client request requires all news services to be contacted in order to be served. However, since news services are contacted in parallel by the collector service, performance degrades better than directly proportional to the number of news services. (Doubling the number of services does not halve the throughput.)
	More interestingly, it seems that the number of clients does not have a strong effect on performance in the interval we considered. To understand this better, in Figure~\ref{fig:scalability_experiment_normalised} we show the result of normalising the data points of each curve in Figure~\ref{fig:scalability_experiment} wrt their maximum speed (in the interval we considered) at 20 services. This gives us an indication of how much performance degrades in percentage wrt the best data point at 20 services, for each client load. We observe that all curves follow similar behaviour, confirming the impression that performance degrades gracefully independently of the number of clients in our interval.
\begin{figure}
\centering
	\begin{tikzpicture}
	\begin{axis}[
		xlabel={News Services},
		ylabel={Throughput (pp/s)},
		legend entries={250 clients, 500 clients, 750 clients, 1000 clients},
		xmin=1,
		xtick=data,
		ytick={100,200,300,400,500,600,700,800,900},
		grid=both
	]
	\addplot table[x=Services,y=250] {results/planet_4cores_services_jolie.dat};
	\addplot table[x=Services,y=500] {results/planet_4cores_services_jolie.dat};
	\addplot table[x=Services,y=750] {results/planet_4cores_services_jolie.dat};
	\addplot table[x=Services,y=1000] {results/planet_4cores_services_jolie.dat};
	\end{axis}
	\end{tikzpicture}
\caption{Scalability experiment.}
\label{fig:scalability_experiment}
\end{figure}
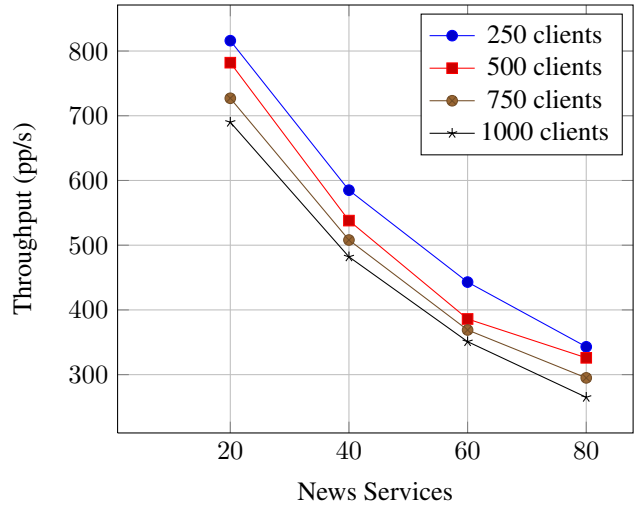
\begin{figure}
\centering
	\begin{tikzpicture}
	\begin{axis}[
		xlabel={News Services},
		ylabel={Throughput (\%)},
		legend entries={250 clients, 500 clients, 750 clients, 1000 clients},
		xmin=1,
		xtick=data,
		ytick={40,50,60,70,80,90,100},
		ylabel near ticks,
		grid=both
	]
	\addplot table[x=Services,y=250] {results/planet_4cores_normalised_jolie.dat};
	\addplot table[x=Services,y=500] {results/planet_4cores_normalised_jolie.dat};
	\addplot table[x=Services,y=750] {results/planet_4cores_normalised_jolie.dat};
	\addplot table[x=Services,y=1000] {results/planet_4cores_normalised_jolie.dat};
	\end{axis}
	\end{tikzpicture}
\caption{Scalability experiment (normalised curves).}
\label{fig:scalability_experiment_normalised}
\end{figure}
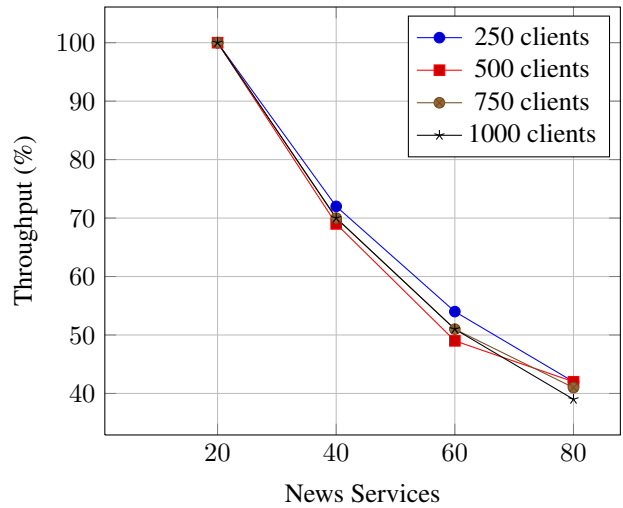
%

%% file: related.tex
%
%
To the best of our knowledge, our work is the first to propose a unified language
for dealing with the programming of web servers, scripting, and the architecture
of service systems in the Web by means of mediator services.
Our development was inspired by related work in these areas, described below.

The frameworks most similar to ours are those for modelling business processes,
such as WS-BPEL~\cite{bpel}, WS-CDL~\cite{wscdl}, and
YAWL~\cite{yawl:journal}.
Differently from our approach, these tools are integrated with web applications
through third-party tools. 
Some of the ideas presented in this paper (e.g., the \lstcode{default}
parameters for implementing web servers) may be easily applied to WS-BPEL,
making our work a potential reference.

The idea of using a router component to obtain RESTful applications is used also in other frameworks, e.g., Ruby on Rails~\cite{rubyonrails}; we can find similar methodologies also in .NET and Java EE. The difference in our approach is that a router is a service. (In general, in Jolie every component in Jolie is a service.) Therefore, it can be independently configured and deployed (or even replicated, if there is a need to scale wrt requests).

Other works offer tools for supporting the development of process-aware web applications.
The papers~\cite{B10,B13} propose a formally-specified language, implemented in Java, for defining 
processes that can transparently access resources on the web using a fixed set of primitive 
operations; the language supports similar process structured as those found in Jolie behaviours,
although in our case operations are user-defined.
In~\cite{DBLP:conf/sac/RossiT08a}, the authors present a
process-based approach to deal with user actions through web interfaces using
EPML; like \Jolie{}, EPML is formally specified and comes with an execution
engine. JOpera comes with an integration
layer for offering REST-based interfaces to business
processes~\cite{DBLP:conf/icsoc/PautassoW11}.
These solutions are formed by integrating separate modules for
process modelling, computation, and system integration. In contrast, our
framework addresses all these aspects using the same language. JOpera also supports the composition of RESTful services using a graphical notation~\cite{P09}.
EPML can integrate with other languages to integrate user
interfaces with process execution; we are currently investigating in a similar direction (see \S~\ref{sec:future}, \emph{Scaffolding of User Interfaces}).
The modelling language IFML~\cite{BF14} captures both processes and the modelling of user front-ends; IFML offers an expressive behavioural model, but is focused on modelling rather than implementation as in here, so it may be an interesting complement to the implementation framework presented in this work.
The S scripting language is a domain-specific language for writing high-performance RESTful web services~\cite{BPPB12}. S is natively based on the Uniform Interface defined in REST and the resource abstraction. This allows for some language-based analyses and optimisations based on the semantics of the Uniform Interface. In our case, we could implement similar features in the router service (\S~\ref{sec:rest_router}), since its configuration tells us which methods are going to be used for accessing which operations/resources. We leave such aspects as an interesting future work, since it may also require an analysis of state modifications as indicated in~\cite{BPPB12}.
In~\cite{P08}, an extension to the BPEL language is proposed to capture the publishing and composition of resources. This is a more high-level approach than that proposed in this paper, where we leave the modelling of resource semantics to the programmer using our more low-level HTTP extension. We see the two approaches as complementary: Jolie may be used as the lower layer of similar high-level abstractions, implemented via a compiler or similar techniques. Indeed, the methodology given in~\cite{P08} should be straightforward to port to a DSL based on our implementation.

Hop~\cite{DBLP:conf/oopsla/SerranoGL06,DBLP:journals/toplas/BoudolLRS12} and
GWT~\cite{gwt} are programming frameworks that deal with the programming of
both the user interface and the server-side application logic using a single
codebase, which gets then compiled in the code for the client interface and the
services.
Differently, in this paper we do not deal with the generation of client code.
Instead, we developed an integration between existing technologies (HTML,
AJAX calls, JSON, etc.) and our services, by using our \lstcode{http} protocol
to convert the data structures handled by these technologies to/from those
handled by Jolie. This leaves the choice of which framework to use for implementing
the web user interface to the developer.
The client code compiled from GWT projects can be reused
with our \lstcode{http} extension, which is able to parse GWT requests.
HipHop~\cite{BS14} is an extension of Hop based on the synchronous
language Esterel~\cite{B00}, which introduces orchestration primitives to Hop.
The major difference between HipHop and our solution is that behavioural code
in Jolie is kept separated from deployment information, making it reusable in 
different environments, whereas HipHop code mixes the two aspects (for example,
cookies in Hop are handled in behavioural code).

Another work that shares some of our aims is the Bigwig project, which offers
a language for the programming of session-aware web applications~\cite{BMS02}.
Our language for behaviours is more expressive than that of Bigwig, which does not
support, e.g., the programming of processes using multiparty sessions; however,
in our setting we obtain this expressiveness by requiring the programmer to manually
handle session identifiers in processes, whereas in Bigwig these are handled automatically.
Bigwig is based on the Apache web server, whereas our approach is
self-contained: web servers, services, and service mediators (which Bigwig does not handle)
are all written in Jolie.

Our \lstcode{default} configuration parameters for \lstcode{http} allows a
service implementation to catch and reply to invocations for operations that
were not known at design time.
The same aspect has been previously theoretically modelled through mobility mechanisms for
names in process calculi, e.g., in~\cite{MPW92,SW01,sock:mobility}.
Our approach is less powerful because these theoretical models elevate the
received operation names at the language level: a service may
receive an operation name, store it in a variable, and then use the latter in
\emph{(input)} and \emph{(output)} primitives as an operation. This is not
possible in our behavioural language, since operations in input and output
statements are statically defined.
We chose not to support this kind of mobility,
since it would make the definitions of \Jolie{} interfaces change at
runtime. This would break the basic assumption of statically defined operations used in
the formal model and implementation of the Jolie language, which goes out of the scope
of this paper.
It would also make Jolie fundamentally different from other
standards for web services, such as WSDL~\cite{wsdl}, with unclear consequences on
their integration.
Static operation names are also used in many formal models for
the verification of concurrent programming languages (e.g., session types~\cite{HVK98}),
which we are interested in adopting for Jolie in the future.

%% file: discussion.tex

We discuss some aspects of web programming with Jolie and future extensions
related to the work that we presented in this article.

%


\paragraph{Holystic Approach}
The main motivation of this work is to lower the complexity of programming
web-based systems by offering a unified language to capture their
different aspects.
However, the current widespread approach of having
a specialised technology for each of such aspects may have
an advantage when it comes to the required knowledge to use them,
as each technology can be studied in isolation.
For example, the administrator of a web server in a larger system
has to learn only how to use the web server
software she uses, abstracting from the other technologies in the rest
of the system (where, e.g., WS-BPEL or ESB technologies may be present).

When dealing with only one aspect of web programming, learning
how to use a specific software to deal with such aspect may
be less time consuming than learning the Jolie language, which
is more general.
A possible solution for this problem could be to develop Domain Specific Languages (DSLs),
supported by Integrated Development Environments (IDEs), that are compiled to
Jolie code. The idea is that a specific DSL would deal with one aspect
of web programming, while retaining the benefits of having a single 
underlying language for the different components of a web system.
It is still uncertain whether this step would really be necessary,
for two reasons.
The first reason is that for simple tasks, such as serving static
content, we can offer a reference distributable implementation such as 
the Leonardo web server in \S~\ref{sec:leonardo}, as an alternative
to other standard implementations such as the Apache Web Server.
The second reason is that many web systems require dealing with
multiple aspects of web programming. In those cases,
it can take less time to learn Jolie than learning about the 
available specific technologies to cover all the use cases that the work
we presented can address; this would amount to learning, at least, a
web server, an orchestration (e.g., WS-BPEL), and a service mediation
technologies.

\paragraph{Adoption}
How and when should a solution such as that proposed in this paper be adopted
in real-world software projects?
We discuss an answer by distinguishing between two main cases: the development of new systems and
the extension of existing systems.

When dealing with the development of an entirely new web system,
Jolie offers a simple and unified language for dealing with 
the architectural aspects (layering, deployment), the behavioural
aspects (application logic), and the serving of static content (web
servers). Therefore, Jolie is now a candidate for the 
rapid prototyping of a web system. Since Jolie integrates with other technologies,
starting with our framework does not imply that the final system
must be written entirely in Jolie: different parts may be refined later
either by using Jolie or other technologies (e.g., Java, WS-BPEL).

When dealing with the extension of an existing system, or even the development
of a new system that has to integrate with other existing systems,
Jolie can be considered as a glue framework for bridging services based on
different technologies. In particular, it is convenient
to use the simple syntax of Jolie for writing processes that direct
the behaviour of other services in a system.
In general, the integration capabilities of Jolie allow for its introduction
in a development team by starting from a single service in a larger system,
which can be used by the team to assess whether Jolie should be adopted in
other parts of the system after seeing how it performs.
We conjecture that this step-by-step introduction of web services written in Jolie
will be key for its adoption by expert web developers.
We are currently following this development methodology in some software projects at the
University of Southern Denmark, for the improvement of the web-based
tools provided to students and staff.

Since Jolie is a relatively new language, most programmers are
still unfamiliar with it and therefore their training must be taken
into account in a project. An advantage of a unified framework
such as ours, though, is that it allows to understand
multiple aspects of web-based systems by learning a single language.
With the rise of more complex and structured web-based systems,
we believe that there can be a motivation to learn Jolie
even for developers who are expert in more established technologies.

%
%
%
%
%
%

\paragraph{Reversibility}
Reversibility techniques (see, e.g.,~\cite{LMS10}) deal with the automatic reversal to previous states in a distributed system. In the context of the Web, reversibility could play a role in allowing users to revert the effects of unsafe operations, e.g., to "un-delete" resources. In complex sessions involving multiple parties, this requires inferring which parties should be notified of such a reversal event. The formal semantics of Jolie~\cite{MC11} should play an important role in enabling the formal study of such notions.

\paragraph{Scaffolding of User Interfaces}
The explicit structure of processes written in Jolie allows us to statically
see the workflow that a user interface should follow when interacting with a Jolie
service.
We could use this aspect to develop a scaffolding tool for user interfaces, starting from the
process structure of a service. Specifically, given a behaviour in
\Jolie{}, it would be possible to automatically generate a user interface that
follows the communication structure of the behaviour.
This would be in line with the notions of \emph{duality} formalised
in~\cite{HVK98,HMS12}.

\paragraph{Behavioural analyses}
Since our framework makes the process logic of a web application explicit,
it would be possible to develop a tool for checking that the invocations
performed by a web user interface written in, e.g., Javascript, match the
structure of their corresponding \Jolie{} service.
The techniques presented in~\cite{HKPYH10,GVRGC10,MY13} may offer useful first
steps towards this aim.

\paragraph{Declarative data validation}
Our framework exploits the message data types declared in the interfaces of
a \Jolie{} service to \emph{validate} the content of incoming messages from
web user interfaces (\S~\ref{sec:type_casting}).
We plan to extend this declarative support for data validation by introducing
an assertion language for message types that can check more complex properties
(e.g., integer ranges and regular expressions).

\paragraph{Extensions to other web protocols}
Our work lays the foundations for using Jolie as a fully-fledged language
to handle HTTP-based systems.
By following the same approach, it would be possible to develop support for
new emerging protocols for the web, such as WebSockets~\cite{websocket} and
SPDY~\cite{spdy}.

%% file: conclusions.tex
We have presented a framework for the programming of process-aware web
systems, where processes are used as a holystic approach to capture the development
of the different components of such systems, such as web servers, orchestrators, and
service mediators.
Through examples, we have shown how our solution subsumes useful
web design patterns and how it captures complex scenarios involving,
e.g., multiparty sessions and evolvability.
Our \lstcode{http} extension is open source and is included in the
standard distribution of \Jolie{}, along with the language additions that we
introduced to support protocol configurations~\cite{jolie:http:source,jolie:website}.
Our integration is seamless wrt data formats, meaning that existing \Jolie{} code can be ported to HTTP without having to  deal explicitly with the typical data formats used in the Web (e.g., JSON).
Since our alterations to the Jolie language targeted only the configuration of communication ports, all the techniques developed for the verification and execution of \Jolie{} programs (as the typing system in~\cite{MC11} for correlation sets)
can be transparently applied to the process-aware web application logic written in our framework.
%